\documentclass[reprint,superscriptaddress,
 amsmath,amssymb,
 aps,
]{revtex4-2}

\usepackage{graphicx}% Include figure files
\usepackage{dcolumn}% Align table columns on decimal point
\usepackage{bm}% bold math
\begin{document}

\preprint{APS/123-QED}

\title{Can Transformers predict system collapse in dynamical systems?}

\author{Zheng-Meng Zhai}
 \affiliation{School of Electrical, Computer and Energy Engineering, Arizona State University, Tempe, AZ 85287, USA}

\author{Celso Grebogi}
\affiliation{Institute for Complex Systems and Mathematical Biology, School of Natural and Computing Sciences, King's College, University of Aberdeen, UK}

\author{Ying-Cheng Lai} \email{Ying-Cheng.Lai@asu.edu}
% \affiliation{School of Electrical, Computer and Energy Engineering, Arizona State University, Tempe, AZ 85287, USA\\
% Department of Physics, Arizona State University, Tempe, Arizona 85287, USA}
\affiliation{School of Electrical, Computer and Energy Engineering, Arizona State University, Tempe, AZ 85287, USA}
\affiliation{Department of Physics, Arizona State University, Tempe, Arizona 85287, USA}

\date{\today}

\begin{abstract}

	Transformer architectures have recently surged as promising solutions for nonlinear dynamical systems, proposed as foundation models capable of zero-shot dynamics reconstruction and forecasting. Despite this success, it remains unclear whether they can truly serve as reliable digital twins of dynamical systems, i.e., whether they capture the underlying physical dynamics in distinct parameter regimes, especially in parameter regimes from which no training data is taken. For parameter-space extrapolation in nonlinear dynamical systems, reservoir computing has demonstrated broad success, as proper training can turn it into an intrinsic dynamical system capable of capturing not only the dynamical climate of the target system but more importantly, how the climate changes with parameter. Transformers, in contrast, rely on permutation-invariant attention mechanisms that can limit their ability to capture how temporal structure changes with parameter. To determine if Transformers have the capability of dynamics extrapolation, we take predicting catastrophic collapse, which occurs when a bifurcation parameter crosses a critical threshold, as a benchmark task. Models are trained on trajectories in normal parameter regimes and then tested on parameters in an unseen regime with system collapse. Our results show that Transformers, across configurations, consistently fail to capture collapse, while reservoir computing reliably predicts the transitions. This surprising finding raises questions about the generalization ability of Transformers to dynamical systems, a topic warranting future research.

\end{abstract}

%\keywords{Suggested keywords}%Use showkeys class option if keyword display desired
\maketitle

%\tableofcontents

% \section{\label{sec:level1}First-level heading:\protect\\ The line
% break was forced \lowercase{via} \textbackslash\textbackslash}

\section{Introduction} \label{sec:intro}

In nonlinear dynamical systems, a catastrophic collapse is typically caused by variations of an underlying bifurcation parameter through a critical threshold. In real-world scenarios, the bifurcation parameters often vary over time and examples of collapse abound, from massive extinction in ecosystems and sudden lake pollution to power grid failures. A common class of critical transisions is crises, where the destruction of a basin boundary leads to the collapse of a chaotic attractor into a chaotic transient~\cite{grebogi1983crises,lai2011transient}. In electrical power systems, voltage collapse can lead to large-scale blackouts, driven by transient chaos near bifurcations~\cite{dhamala1999controlling}. An example of significant interest is the Atlantic Meridional Overturning Circulation, where a shift in hidden parameters can induce a state flip that disrupts the warm climate of Europe~\cite{zhai2024machine}, with a saddle-node bifurcation as the responsible dynamical mechanism.

Often, accurate governing equations are rarely available in practice, while time-series data are ubiquitous. In light of this, machine learning-based digital twins are increasingly used as surrogates for target dynamical systems, enabling simulations of parameter changes, control, and inverse modeling~\cite{chakraborty2021machine}. Recent empirical studies~\cite{KFGL:2021a,patel2021using,kong2023reservoir,PKMZGHL:2024,yan2024emerging,zhuge2025deep} have demonstrated that digital twins can predict critical transitions or tipping points in nonlinear dynamical systems. These approaches typically involve training models on system trajectories from safe parameter regimes, and then testing on trajectories generated under unseen parameter values into the post-critical regime. The parameter values are injected into the neurons in the network through a parameter channel. In this regard, reservoir computing~\cite{Jaeger:2001,jaeger2004harnessing} has been especially successful~\cite{PLHGO:2017,LPHGBO:2017,PHGLO:2018}: as a dynamical system in itself, it can naturally imitate the target system and adapt to out-of-domain regimes. A potential perspective for interpreting reservoir computing is that its dynamics are generally an embedding of the target dynamical system, which can be deemed as generalized synchronization~\cite{hart2020embedding,gauthier2021next,PC:2025}. However, the finite network size in a reservoir computer constrains its predictive capability. (A brief description of related works of exploiting machine learning for dynamical systems and anticipating critical transitions can be found in Appendix~\ref{appendix:related_works}.)

The past few years have witnessed the remarkable success of the Transformer architecture, the most influential sequence model to date~\cite{vaswani2017attention}. Initially dominating natural language processing, Transformers have since percolated into fields such as time-series forecasting~\cite{SZLWZH:2025}, video analysis~\cite{SJENMC:2023}, speech processing~\cite{liu2021tera}, and increasingly, nonlinear dynamics~\cite{zhai2025bridging}. One key advantage of Transformers is their generalization ability, particularly in out-of-domain tasks, as seen in phenomena such as ``grokking'' in large language models~\cite{PBEBM:2022}. In nonlinear dynamics, Transformers have been applied to reconstruct unseen systems from sparse or random observations~\cite{zhai2025bridging}, and time-series based large models have even been applied or built for zero-shot forecasting of new dynamics~\cite{zhang2024zero,LBG:2025}. Yet, no theoretical guarantee exists that a trained Transformer can perform as a faithful digital twin, either for the systems it was trained on or for unseen test systems. Indeed, a study~\cite{zhang2025context} suggests that even large pretrained models such as Chronos underperform simple baselines such as parroting. In mechanisms, this may not be surprising: reservoir computing resembles the target system by being a dynamical system itself, while the Transformer relies on permutation-invariant self-attention, which is less sensitive to temporal ordering~\cite{zeng2023transformers}. This raises a fundamental question: Can Transformers serve as digital twins of dynamical systems?

To address this question, we focus on a classical yet significant challenge: machine-learning prediction of critical transition and tipping. In chaotic systems, short-term prediction is meaningful only up to several cycles of natural oscillations (or equivalently, several Lyapunov times), beyond which the predicted trajectory diverges from the ground truth. In comparison, the ability to capture critical transitions under changing parameters becomes a decisive test of the ability of the model as a digital twin. Following prior works~\cite{KFGL:2021a,KFGL:2021b} on reservoir-computing based prediction of critical transitions, we aim to study systematically if Transformers are capable of anticipating critical transitions.

The results are unexpected. Transformers train effectively on trajectories within safe parameter regimes, achieving strong multi-step prediction performance. However, when tested on unseen parameter values corresponding to collapse states, Transformers fail to capture the transition. For example, instead of predicting a state shift, they produce persistent oscillations as in the training regime. We test four representative systems: a chaotic food-chain system, a power system, the Ikeda map, and the Kuramoto-Sivashinsky equation, which vary in dimension and complexity. Despite extensive attempts to mitigate overfitting, such as broadening the training parameter regimes or adjusting the number of parameters in Transformer, none of our Transformer variants successfully predicted collapse in any system. These findings indicate that a critical rethink of the role of Transformers in nonlinear dynamics may be necessary. 

%To our knowledge, this is the first work to systematically challenge the effectiveness of Transformers as digital twins in dynamical systems for anticipating system collapse. We conduct comprehensive experiments across multiple systems and transformer settings, and benchmark against established solutions. While reservoir computing reliably succeeds at this task, standard Transformer configurations consistently fail. We propose critical transition and tipping prediction as a benchmark task for evaluating digital twins, complementing standard forecasting metrics, and highlight implications for future research.

Our study suggests that the widely reported generalization ability of Transformers may be overstated, at least in the context of serving as digital twins for predicting critical transitions and tipping points. By contrast, reservoir computing remains highly effective, although its scalability limits prevent it from functioning as a general foundation model. Looking ahead, we hope this work stimulates renewed attention to physics-guided architectures, and inspires the development of next-generation machine learning frameworks that can faithfully capture not only the state evolution of dynamical systems, but also how their behavior shifts under parameter variation, thereby serving as reliable digital twins.

\section{Methods} \label{sec:method}

\begin{figure*} [ht!]
\begin{center}
\includegraphics[width=\linewidth]{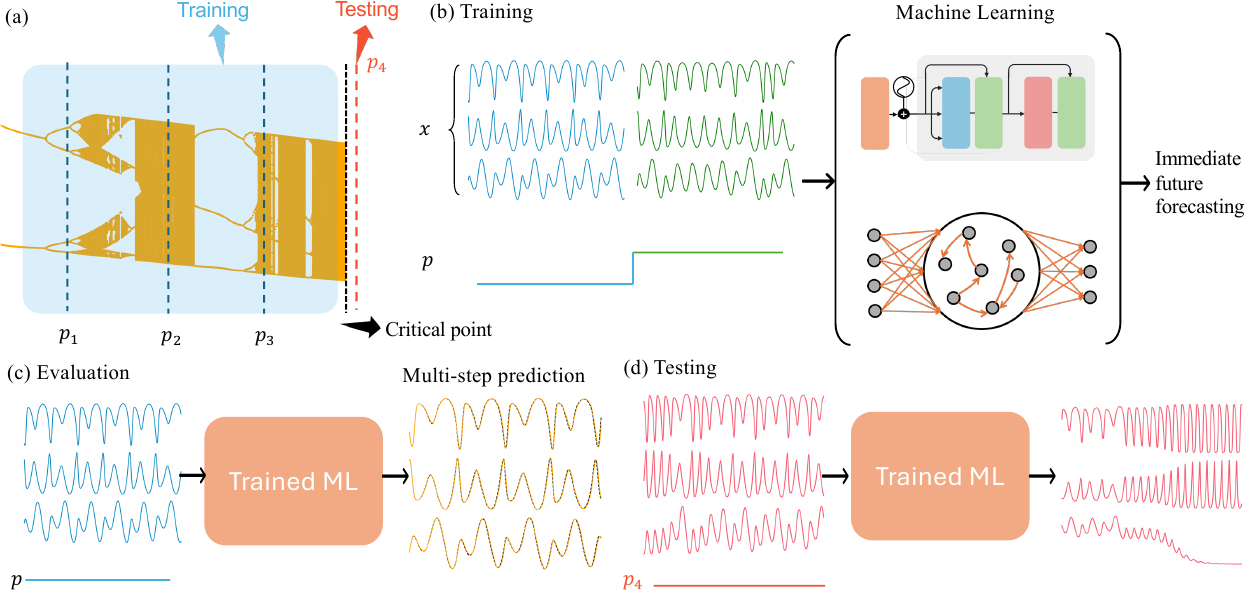}
\end{center}
\caption{Parameter-adaptable machine-learning framework for anticipating critical transitions. (a) A schematic bifurcation diagram. As the bifurcation parameter $p$ varies, the system remains in a safe regime (periodic/chaotic attractors) until a critical point $p_c$ (black vertical line), beyond which trajectories exhibit transient chaos followed by collapse. (b) Training: the machine-learning model is trained on trajectories from safe parameters $p < p_c$, with the input formed by the time series and the parameter channel. (c) Trained models are evaluated by multi-step predictions at training parameters. (d) Testing. Given a new unseen parameter $p_4 > p_c$, the model is tasked with forecasting forward. A successful prediction reproduces the collapse (transient followed by decay), rather than sustained oscillations.}
\label{fig:overview}
\end{figure*}

\subsection{Parameter-adaptable machine learning for predicting critical transitions}

We motivate and introduce parameter-adaptable machine learning for predicting critical transitions and tipping points. The overview of the framework is illustrated in Fig.~\ref{fig:overview}. In particular, Fig.~\ref{fig:overview}(a) shows a schematic bifurcation diagram of a typical nonlinear system: as the bifurcation parameter varies, the qualitative behavior of the system changes accordingly. For any parameter value in the orange region (covered by the blue window), the asymptotic attractor can be chaotic or periodic; we refer to this as the safe regime. At a critical parameter value marked by the vertical black line, a catastrophic bifurcation, e.g., a crisis, occurs, which destroys the chaotic attractor, after which trajectories exhibit a chaotic transient followed by collapse. The generic feature is that, to the left of the critical point $p_c$, the time series remain oscillatory (periodic or chaotic), whereas slightly to the right of $p_c$ the dynamics transition to collapse after a transient. It has been demonstrated~\cite{KFGL:2021a,KFGL:2021b,PKMZGHL:2024,PKGHL:2026} that, even when the system is currently in a safe regime, adaptable reservoir computing is capable of anticipating the collapse and giving accurate prediction of the critical parameter value or the transition point. The question is whether Transformer possesses the same predictive capability.

We consider the previously justified setting~\cite{KFGL:2021a,KFGL:2021b,PKMZGHL:2024,PKGHL:2026} for predicting critical transitions with reservoir computing, in which only a few, e.g., three, parameter values in the safe regime are known, together with the corresponding time series that are long enough to ensure satisfactory training and testing performances. More specifically, let $p_1$, $p_2$, and $p_3$ denote the three parameter values. The goal is to train a machine-learning model such that, when presented with a new parameter $p_4 > p_c$, it correctly anticipates collapse. The training protocol is shown in Fig.~\ref{fig:overview}(b), which follows a parameter-adaptable design~\cite{KFGL:2021a}. Let $\mathbf{x}$ denote the system state time series and $\mathbf{p}$ be the associated parameter channel. The model input is the concatenation $[\mathbf{x};\,\mathbf{p}]$. During training, a one-step-ahead prediction loss is minimized. However, this single-step accuracy is often insufficient to certify learning in chaotic systems, so we further validate the trained model via multi-step forecasting under the same training parameters, as shown in Fig.~\ref{fig:overview}(c). A machine-learning model is deemed well trained if it can produce accurate multi-step predictions over several oscillation cycles for the three chosen parameter values in the safe regime.

The performance of the model on the critical-transition prediction task can be assessed by providing the test parameter $p_4 > p_c$ and asking the model to forecast forward from a random initial state, as depicted in Fig.~\ref{fig:overview}(d). A successful prediction is one in which the model correctly anticipates collapse (i.e., reproduces a transient followed by decay), rather than persisting in an oscillatory state. 

In what follows, we apply this evaluation protocol to three benchmark chaotic systems and compare two parameter-adaptable architectures: reservoir computing and Transformer, under multiple neural-network configurations. (A detailed description of the architectures of Transformer and reservoir computing is provided in Appendix~\ref{appendix:ML_Methods}.) It is worth noting that, the generalization task here differs fundamentally from tasks in natural language processing, as it requires sensitivity to parameter variation and stability thresholds, which may raise challenge for self-attention mechanisms.

\subsection{Computational setting}

{\em Datasets}. The benchmark includes a three-dimensional food-chain system, a four-dimensional voltage system, and the discrete Ikeda optical-cavity map. These systems do not satisfy sparsity conditions, and thus their governing equations cannot be identified by sparse regression methods (e.g., SINDy-type methods)~\cite{WYLKG:2011,BPK:2016,Lai:2021}. Among them, the Ikeda map is the most difficult case, and yet no existing method can faithfully recover its governing equations from data. In addition, we also include the Kuramoto-Sivashinsky system as a representative spatiotemporal chaotic system. More details about these systems can be found in Appendix~\ref{appendix:results}. By varying system parameters, each system can be driven from a stable to an unstable (collapsed) state.

{\em Machine-learning models}. We adopt parameter-adaptable reservoir computing and Transformer, where chaotic time series and the corresponding parameters are combined as model inputs. For reservoir computing, we follow the same simulation setup as in Refs.~\cite{KFGL:2021a,KFGL:2021b}, with network sizes typically smaller than 1000. A distinctive feature of reservoir computing is that the input weights and the network adjacency matrix remain fixed during training, and only the output weights are optimized. Under this scheme, the number of trainable parameters is on the order of $10^3$. For the Transformer, we use a decoder-only autoregressive model. In the numerical experiments, we set the maximum input length $L_{\max} = 512$ and use order of $10^6$ trainable parameters. Detailed reservoir computing and Transformer configurations are described in Appendix~\ref{appendix:ML_Methods}.  

{\em Pipelines}. To evaluate the ability of each machine-learning model to predict critical transitions, we generate 500,000 data points for each bifurcation parameter. For the continuous systems, one data point corresponds to approximately 1/50 of a cycle of oscillation. The task proceeds as follows: We first train the model under safe parameter regimes until it can achieve multi-step prediction performance comparable to the ground truth. We then test the model on unseen parameter values expected to induce collapse. A successful prediction is defined as correctly forecasting the state shift (collapse). In our experiments, for training each system, we provide three specific parameter values and the corresponding chaotic time series. The training dataset size differs substantially between reservoir computing and Transformer: the former typically requires about $10^3\sim10^4$ points, whereas Transformers require about $10^6$. In both cases, we use the minimal training size sufficient for accurate multi-step prediction in the training regime. To ensure statistical reliability, we repeat each experiment 50 times independently for each system. All training experiments were conducted on two servers, each equipped with 6 NVIDIA RTX A6000 GPUs.

{\em Metrics}. We evaluate multi-step forecasting accuracy using the root mean square error (RMSE):
\begin{align}
\text{RMSE}(X, \hat{X}) = \sqrt{\frac{1}{T_{\text{p}} d}\sum_{t=1}^{T_{\text{p}}}\sum_{j=1}^{d} \bigl(X_{t,j} - \hat{X}_{t,j}\bigr)^2},
\end{align}
where $X \in \mathbb{R}^{T_{\text{p}} \times d}$ and $\hat{X} \in \mathbb{R}^{T_{\text{p}} \times d}$ denote the ground-truth and predicted trajectories, $T_{\text{p}}$ is the prediction horizon, and $d$ is the system dimension. 

To assess the performance in predicting critical transitions, we use the collapse prediction rate $P_c$. Within a search range of supercritical bifurcation parameters $p>p_c$, a trial is counted as ``collapse'' if the predicted trajectory falls below a predefined threshold $\theta$ and remains there for a sufficient duration. Formally,
\begin{align}
P_c = \frac{1}{M}\sum_{m=1}^{M}\mathbf{1}\{\text{collapse detected in trial} \; m\},
\end{align}
where $M$ is the number of independent trials (e.g., model seeds or warm-up segments).

\section{Results} \label{sec:Results}

For each benchmark system, we generate trajectories from three distinct bifurcation-parameter values within the chaotic or periodic regime for training. The task is to predict dynamics at an unseen parameter value that induces collapse. A successful prediction is not the continuation of a periodic or chaotic attractor but rather a collapse, i.e., a sudden drop of the system state to a lower constant value. Additional experiments exploring Transformer architectures and training settings can be found in Appendix~\ref{appendix:further_transformer}.

\subsection{Critical transition prediction in the food-chain system}

\begin{figure*} [ht!]
\centering
\includegraphics[width=\linewidth]{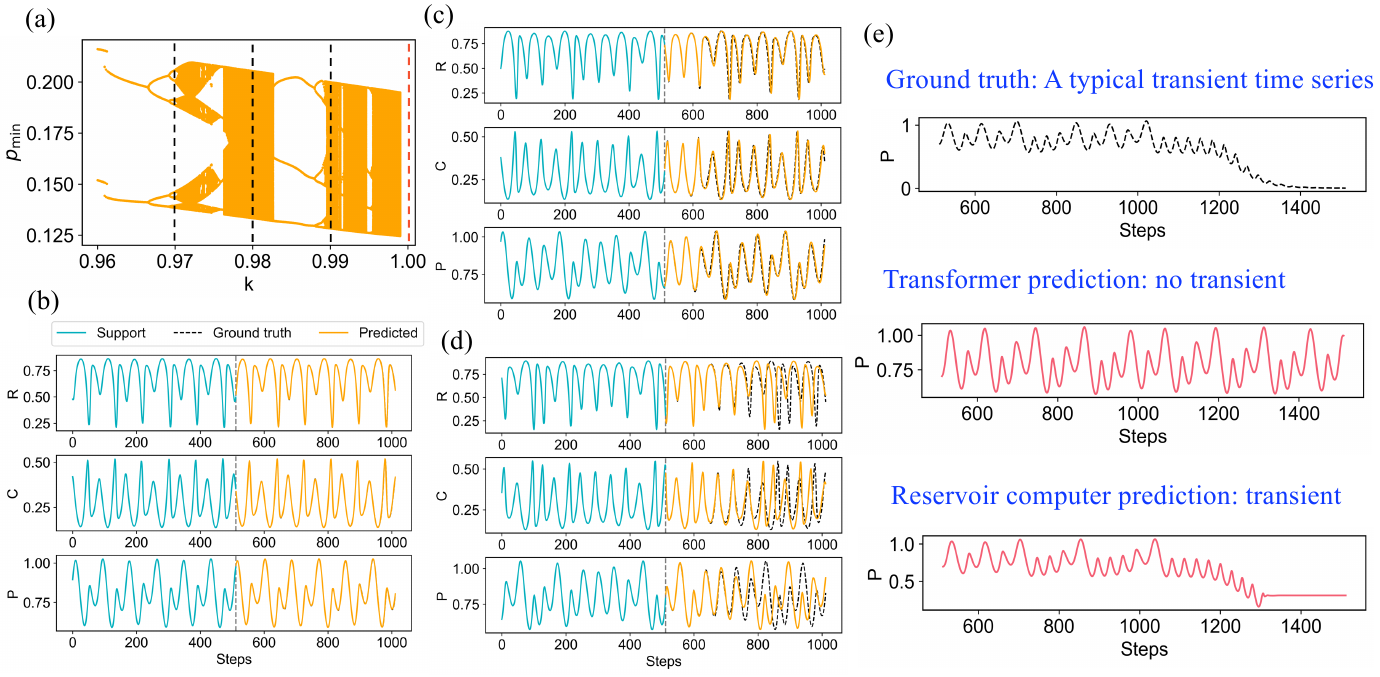}
\caption{Predicting a critical transition in the chaotic food-chain system. (a) Bifurcation diagram, with training parameters (black dashed) and the testing parameter beyond the critical point (red dashed). (b-d) Multi-step predictions by the Transformer for $K=0.97, 0.98,$ and $0.99$, respectively. Blue curves indicate warm-up input, and orange curves denote closed-loop Transformer predictions. (e) Comparison of Transformer and reservoir-computing predictions at $K=1.0 > K_c$. The ground truth trajectory collapses after a transient; the reservoir computer correctly reproduces this collapse, whereas the Transformer continues oscillating.}
\label{fig:foodchain}
\end{figure*}

For clarity, we first present detailed results for the chaotic food-chain system, while deferring the results from the other benchmark systems to Appendix~\ref{appendix:results}. The food-chain system provides a canonical ecological example where a parameter drift can cause catastrophic collapse. It models the interactions among three species: resources $R$, consumers $C$, and predators $P$, governed by
\begin{align}
\frac{dR}{dt} &= R \left(1 - \frac{R}{K}\right) - \frac{x_c y_c C R}{R + R_0}, \nonumber \\
\frac{dC}{dt} &= x_c C \left(\frac{y_c R}{R + R_0} - 1\right) - \frac{x_p y_p P C}{C + C_0}, \\
\frac{dP}{dt} &= x_p P \left(\frac{y_p C}{C + C_0} - 1\right), \nonumber
\end{align}
where $R$, $C$, and $P$ are the population densities of the three species. The bifurcation parameter $K$ controls the environmental carrying capacity of the resource. As $K$ increases, the system undergoes a boundary crisis bifurcation, as illustrated in Fig.~\ref{fig:foodchain}(a), with a critical point near $K_c \approx 0.99976$. For $K < K_c$, the system exhibits sustained chaos; for $K > K_c$, chaotic transients eventually collapse to extinction. Other constants ($x_c, y_c, x_p, y_p, R_0, C_0$) are chosen to be ecologically reasonable and are remained fixed. 

We train both models on chaotic trajectories at $K=[0.97, 0.98, 0.99]$, all within the safe regime, as illustrated in Fig.~\ref{fig:foodchain}(a). During training, both reservoir computing and Transformer achieve accurate multi-step forecasting over several Lyapunov times. The performance of Transformer in this stage is shown in Figs.~\ref{fig:foodchain}(b-d), where it closely tracks the true dynamics. Afterwards, we test the models at a parameter beyond the critical point ($K=1.0 > K_c$). In the ground truth, the predator density $P$ undergoes transient chaos followed by collapse. Figure~\ref{fig:foodchain}(e) shows the failure of the Transformer, as it continues generating oscillatory trajectories resembling those in the training regime, and never predicting collapse. By contrast, the reservoir computer initially follows the transient oscillations and subsequently forecasts the collapse, consistent with the ground truth. In this setting, none of our experiments with Transformers resulted in success, despite extensive tuning of architectures and training ranges. 

\subsection{Predicting long-term attractor and critical transition}

\begin{figure*}[ht!]
\centering
\includegraphics[width=\linewidth]{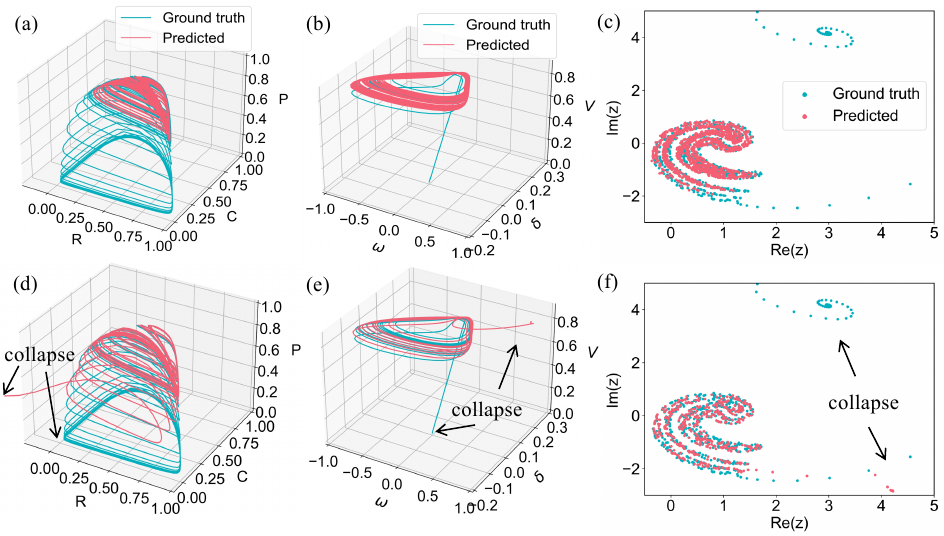}
\caption{Long-term predictions across benchmark systems. Top and bottom rows are for Transformer and reservoir-computing predictions, respectively. (a, d) Food-chain system. (b, e) Power system. (c, f) Ikeda map. Ground-truth dynamics undergo collapse beyond the critical bifurcation parameter, which is faithfully reproduced by the reservoir computer but consistently missed by the Transformer.}
\label{fig:attractor}
\end{figure*}

An important consideration is that the predictions in Fig.~\ref{fig:foodchain} are shown only over relatively short horizons. In practice, machine-learning models may exhibit different reaction times to parameter changes. For $K \agt K_c$, the deviation from the critical point may be small, and the model may not respond immediately. Indeed, we observed that across different trained reservoir computers, and even across different warm-up segments for the same model, the predicted collapse occurred at distinct times under the same parameter. It is worth emphasizing that, no method can predict the exact timing or the post-critical state as the collapse is assumed to occur in the future and no information about the collapsed state has been fed into the machine-learning model. At best, a well-trained model can indicate that a transition will occur within a future time window. This perspective is necessary for interpreting collapse prediction: one may ask whether the Transformer's apparent ``failure'' in Fig.~\ref{fig:foodchain} might eventually collapse given a longer horizon, or conversely, whether the reservoir computer's apparent success might eventually revert to oscillations. Addressing these questions requires the machine-learning model to generate the long-term attractor of the target system.

To evaluate long-term behavior, we extend the predictions of both Transformer and reservoir computing under the same experimental settings as before and present results for three benchmark chaotic systems here, while detailed system descriptions are provided in Appendix~\ref{appendix:results}. A brief description of the two benchmark systems other than the food-chain system are as follows. The power system is a four-dimensional electrical system in which chaotic fluctuations can cause voltage collapse as the load parameter $Q_1$ varies. We train on three safe chaotic parameters $Q_1 = [2.98968, 2.98973, 2.98978]$ and test at $Q_1 = 2.989830 > Q_{1c}$. The Ikeda map, in contrast, is a two-dimensional nonlinear map describing the dynamics of a laser pulse in a nonlinear cavity, with the dimensionless input amplitude $\mu$ as its bifurcation parameter. Training is performed at $\mu = [0.91, 0.94, 0.97]$, and testing at $\mu = 1.01 > \mu_c$.

Examples of long-term predictions are shown in Fig.~\ref{fig:attractor}. The first row presents Transformer predictions, and the second row shows the reservoir-computing predictions, with the three columns corresponding to the food-chain system, the power system, and the Ikeda map, respectively. For the power system, which has four dynamical variables, we visualize the last three; for the Ikeda map, which involves a single complex variable $z$, we plot its real and imaginary components. For fair comparison, both predictions of the two models are displayed on the same scale within each system. The results reveal a consistent pattern: the Transformer persistently generates stable attractors and never transitions to collapse, even when the ground truth does. In contrast, the reservoir computer successfully predicts collapse in all cases, as illustrated in panels (d–f). We note that in the power system, the ground truth collapse involves a sharp excursion to an extreme value before stabilizing at a constant. For visualization clarity, this extreme value is omitted.

\begin{table*}
	\caption{ \label{tab:performance} Performance on critical transition prediction across systems. The notion Transformer$_b$ stands for parameter-adaptable Transformer architexture with a bias (see Appendix~\ref{appendix:ML_Methods}).}
\begin{ruledtabular}
\begin{tabular}{ccccccc}
\textbf{Target system} & \textbf{ML Model} & \textbf{RMSE}$\downarrow$ & 
$P_c(\text{collapse})\uparrow$ & \textbf{Pred.\ CT} & \textbf{GT CT} & \textbf{$T_{\rm train}$[s]}$\downarrow$\\ \hline
\\
Food chain & Transformer & 0.026 & 0.38  & 1.14958   & 0.99976   & 5744 \\
           & Transformer$_b$ &  0.020 &  0.00 &  N/A & 0.99976   & 5760 \\   
           & Reservoir Computer  & 0.002 & 1.00  & 0.99986   & 0.99976   & 0.67 \\
\\
Voltage    & Transformer & 0.037 & 0.001 & 3.02578   & 2.98983   & 5095 \\
           & Transformer$_b$ & 0.048 & 0.02 & 3.03208 & 2.98983  & 5080 \\
           & Reservoir Computer & 0.004 & 1.00  & 2.98983   & 2.98983   & 1.4 \\
\\
Ikeda map  & Transformer & 0.219 & 0.00  & N/A       & 1.0027    & 5100 \\
           & Transformer$_b$ & 0.370 &  0.00 & N/A & 1.0027    &  5104 \\
           & Reservoir Computer & 0.055 & 1.00  & 1.0073    & 1.0027    & 0.06 \\
\end{tabular}
\end{ruledtabular}
\end{table*}

\subsection{Statistics of critical transition prediction}

The examples presented above are representative of the general behavior observed across our experiments, rather than isolated cases. To place this conclusion on a more solid foundation, we present statistical evaluations. Across all trials, Transformers never predicted a single collapse, regardless of configuration, even though their multi-step predictions in the training regime confirmed successful learning. Reservoir computing, in contrast, was capable of predicting collapse, though not deterministically: in few trials it produced oscillatory trajectories instead. Such variability is expected, as parameters only slightly above the critical threshold may not elicit sufficiently sensitive responses from every trained model.

A natural assumption is that a sufficiently expressive and well-trained model, when provided with a bifurcation parameter $p > p_c$, should eventually predict collapse if $p$ is gradually increased. To formalize this, we adopt the following criterion: starting from $p$ slightly above $p_c$ and increasing the parameter step by step. For each value, the trained model generates a predicted trajectory. If the trajectory remains oscillatory below a certain $p$ value but collapses beyond it, we define the corresponding parameter as the predicted critical point of that model. Based on this criterion, we conduct statistical experiments for both Transformers and reservoir computing. Specifically, we independently train 50 models for each architecture. For each trained model, we generate 20 predictions with different warm-up segments, yielding $50 \times 20 = 1000$ simulations in total. We then record, within a predefined search range, the fraction of cases in which the model predicts collapse. We denote this probability as $P_c(\text{collapse})$. Notably, it is computed regardless of the predicted critical value itself, i.e., Transformer may predict collapse only at a parameter much larger than the ground-truth $p_c$, but such cases are still counted insofar as collapse occurs. It is worth noting that the search ranges differ between architectures: Transformers, being less sensitive to parameter variations, are evaluated over a broader interval, whereas reservoir computing typically achieves near $100\%$ collapse prediction within a narrower range. Furthermore, we define collapse strictly as a dynamical drop of the observable variable below its oscillatory range. In cases with large bifurcation parameters, the system may instead converge directly to a constant state without such a drop. These cases are not counted as successful collapse predictions.

The results are summarized in Tab.~\ref{tab:performance}, which compares short-term forecasting accuracy in training and collapse prediction performance in testing across the three benchmark systems. RMSEs are reported as the mean error across the three training parameters, with multi-step forecasting horizons of 250, 250, and 10 steps for the food-chain system, power system, and Ikeda map, respectively, corresponding to approximately $4\sim5$ Lyapunov times. Transformer$_b$ denotes the Transformer variant that includes the same parameter-bias term used in reservoir computing (see Appendix~\ref{appendix:PARC_Transformer} for details). While Transformers require nearly two orders of magnitude more training data and more than 5000 seconds of training time, their RMSE values do not match the efficiency or accuracy of reservoir computing. Nevertheless, as illustrated in the examples, the multi-steps predictions by Transformer are reasonable, as maintaining accuracy for $4\sim5$ Lyapunov times is generally regarded as strong forecasting performance for chaotic systems. In comparison, collapse prediction exposes a sharp divergence in performance. The reservoir computer consistently achieves $100\%$ success and highly accurate estimates of the critical point, while Transformers fail: the collapse prediction probabilities are only $0.38$, $0.002$, and $0.0$ for the three systems. Moreover, even in the rare cases where Transformers do predict collapse, the estimated critical points deviate substantially from the ground truth. This discrepancy is nontrivial. For example, as shown in Fig.~\ref{fig:foodchain}(a), shifting the bifurcation parameter $K$ from $0.97$ to $0.98$ already traverses qualitatively different dynamics, so even seemingly small numerical errors correspond to severe dynamical mismatches. Further statistical analysis are provided in Appendix~\ref{appendix:add_statistics}.

\section{Heuristic mechanistic analysis of failure of Transformer in anticipating critical transitions} \label{sec:theory}

We provide a heuristic comparison between parameter-adaptable reservoir computing and Transformers in the context of constructing a digital twin for nonlinear dynamical systems.

Reservoir computing relies on its hidden reservoir state $r(t)$, which evolves according to Eq.~(\ref{eq:rc}), based on both its previous state and the current input. This makes RC itself a driven dynamical system, excited by the target system state and parameter $(x(t),p(t))$. Under the echo-state property, the driven reservoir converges to a unique trajectory determined by the input history~\cite{jaeger2007optimization}. In this regime, the reservoir establishes a form of generalized synchronization: $r(t)$ becomes an approximate high-dimensional embedding of the underlying attractor for each parameter value.

Importantly, after training, a reservoir computer does not simply integrate information within a finite history window. Instead, it evolves autonomously, on a learned manifold with $p$ acting as a continuous control input under the parameter-aware design. Small variations in $p$ therefore move the driven reservoir state smoothly along a high-dimensional manifold that mirrors how the physical attractor deforms with parameter. This induces a strong inductive bias: when a boundary crisis occurs and the physical attractor disappears, the reservoir asymptotic dynamics naturally change as well. Thus, even without being trained directly ``on the bifurcation,'' the reservoir computer inherits sensitivity to global stability structure and can fall off the embedded chaotic manifold into the collapsed state for the corresponding parameter.

In contrast, a vanilla Transformer with causal attention implements a static mapping $(x,p)_{t-L+1:t} \mapsto \hat{x}_{t+1}$, where history enters only through self-attention over a finite window of length $L_{\max}$. Self-attention is fundamentally permutation-invariant. Although positional encodings reintroduce temporal order, the architecture remains a feed-forward function from a finite sequence to the next step; it lacks an internal state governed by a dynamical law, and thus do not change the fact that self-attention lacks built-in notions of stability, phase space, or bifurcations. As a result, Transformers primarily learn sophisticated autoregressive rules: mappings from recent oscillatory patterns to the next value. This is sufficient to reproduce accurate short-term forecasts, and even approximate the invariant measure of the attractor within the training (safe) regime, but it does not require reconstruction of a global state space or vector field. Consequently, the parameter channel $p$ could be often learned only as a mild modulation of amplitude or frequency, rather than as a control capable of shifting or destroying attractors.

Critical transitions, however, are governed by global bifurcations such as boundary crises, which depend sensitively on the geometry of stable and unstable manifolds. Capturing them requires, at least approximately, embedding the basin boundary and tracking how it moves with $p$. RC naturally acquires this through generalized synchronization and the echo-state property. Transformers, despite powerful pattern-extraction ability, lack mechanisms to encode unstable manifolds or Lyapunov structure. They fit local temporal correlations on the attractor but fails in learning how global stability depends on $p$. As a result, for $p>p_c$, the Transformer closed-loop dynamics tend to remain locked onto an ``effective'' attractor inherited from the safe regime, even though the true attractor has already disappeared.

\section{Discussion} \label{sec:discussion}

Transformers have achieved remarkable success in natural language processing, computer vision, and related domains, and have recently been proposed as foundation models for nonlinear dynamics. This work questions the effectiveness in this new setting specifically for predicting nonlinear dynamical systems. Rather than focusing on standard trajectory forecasting, we pose a challenge that is real-world important: critical transition and tipping prediction. The task requires models trained in regimes of periodic or chaotic oscillations, to predict under unseen parameters, that the system should collapse to extinction rather than maintain oscillations. To provide a baseline, we systematically compare Transformers with reservoir computing across benchmark low-dimensional and high-dimensional chaotic systems. 

The results are striking. While both Transformers and reservoir computing achieve accurate multi-step forecasting in the safe parameter regime, their generalization diverges sharply under parameter drift. Reservoir computing consistently anticipates collapse and locates the critical point with high precision. Whereas Transformers, across all tested configurations, typically fail to predict collapse. In the rare cases where Transformers succeed, the prediction arises only when the injected parameter value is excessively large. This raises a fundamental question: can Transformers truly serve as faithful digital twins of dynamical systems? From a physics perspective, the performance gap stems from architectural principles. Transformers, relying on permutation-invariant self-attention, excel at sequence pattern extraction but lack inductive bias toward parameter-induced regime changes. In contrast, reservoir computing, being a dynamical system itself, naturally embeds target dynamics and generalizes across bifurcations. 

We stress that our findings do not lie in advocating reservoir computing as the ultimate solution for nonlinear dynamical systems. Indeed, a reservoir computer itself is limited by finite reservoir size and scalability constraints. Instead, our goal is to highlight an important and surprising comparison, raise the question of whether Transformers capture dynamical principles, and suggest possible mechanisms behind their failure.

Beyond the results presented, prior studies show that reservoir computers can reconstruct entire bifurcation diagrams, including transitions between periodicity and chaos, by training on a few parameter values~\cite{kong2023reservoir}. This explains strong extrapolation ability of reservoir computing in our experiments: it learns the underlying relationship between parameter and dynamics, rather than just trajectories. These observations also suggest examining whether Transformers can similarly recover bifurcation structures within the safe regime, as a valuable direction. In addition, several related challenges remain open, such as building data efficient digital twins under limited observations~\cite{zhai2025learning}, and predicting not only the onset of collapse but also the post-collapse equilibrium. Incorporating physical constraints or structural priors may help address these gaps.

In language modeling, training defines a probability distribution over token sequences. In-distribution generalization represents generating new samples from this distribution, whereas out-of-distribution generalization corresponds to producing samples from a new distribution, given a few examples as context. In nonlinear dynamics, training on the pre-collapse attractor encourages a Transformer to over-index on this specific distribution. Recent mechanistic studies show that Transformers trained on a variety of synthetic sequence tasks often implement, or in-context learn, finite- or variable order Markov models~\cite{ildiz2024self,edelman2024evolution,chen2024unveiling}. This suggests that Transformers may treat dynamical attractors as fixed sequence distributions, rather than reconstructing the underlying flow or its parameter dependence. (see Sec.~\ref{sec:theory} for a more detailed heuristic mechanistic discussion.)

Looking forward, several avenues merit exploration. One direction is the design of hybrid models, such as reservoir-attention architectures~\cite{koster2024attention}, which combine the dynamical embedding ability of reservoir computing with the scalability of Transformers to incorporate parameter-state relationships. Another is scaling: whether sufficiently large models, trained with diverse parameter regimes and domain-specific objectives, can eventually succeed remains an open question. Although foundation models do not perform well in system collapse prediction under zero-shot settings (see Appendix~\ref{appendix:foundation}), fine-tuning could possibly improve their performance. Exploring alternative parameter-conditioning mechanisms for Transformers is also important; variants such as physics-informed~\cite{zhao2023pinnsformer}, continuous-time~\cite{chen2023contiformer}, or state-space~\cite{gu2024mamba} Transformers may offer useful inductive biases. More broadly, physics-informed architectures and bifurcation-aware priors may be essential for robust digital twins of nonlinear systems. Reliable digital twins of nonlinear dynamics will play a significant role in forecasting critical events in ecology, climate, power grids, and beyond. We hope this work will stimulate renewed attention to physics-guided architectures and inspire the development of next-generation machine learning frameworks that can faithfully capture not only the state evolution of a dynamical system but also how its behavior shifts, especially for critical transitions and tipping.

\section*{Code availability}

The codes for generating all the results can be found on GitHub: https://github.com/Zheng-Meng/RC-Transformer-Tipping-Point.

\begin{acknowledgments}
This work was supported by the Army Research Office under Grant No.~W911NF-26-2-A002 and by the Office of Naval Research under Grant No.~N00014-24-1-2548.
% the Air Force Office of Scientific Research under Grant No.~FA9550-21-1-0438 and by the Army Research Office under Grant W911NF-24-2-0228.
\end{acknowledgments}

\appendix

\section{Related works} \label{appendix:related_works}

\subsection{Machine learning for dynamical systems}

Recurrent neural networks (RNNs) are designed to capture temporal dependencies in sequential data. However, standard RNNs are difficult to train on chaotic time series~\cite{mikhaeil2022difficulty}. Gated architectures such as LSTM (Long Short-Term Memory) networks or GRUs (Gated Recurrent Units)~\cite{vlachas2020backpropagation}, with strong stability constraints, can mitigate this issue by ``locking'' the network into stable dynamics. More recently, a modified teacher forcing scheme for RNNs was proposed~\cite{hess2023generalized}, which strictly bounds gradients in chaotic regimes, enabling faithful attractor reconstruction. In another direction, it has been found that reservoir computing, also a class of RNNs, provides high-performance, model-free prediction of chaotic systems at low cost~\cite{PHGLO:2018}. Since its introduction~\cite{Jaeger:2001,jaeger2004harnessing} and the demonstration of its ability to predict spatiotemporal chaotic systems~\cite{PLHGO:2017,LPHGBO:2017,PHGLO:2018}, reservoir computing has stimulated both theoretical and applied developments~\cite{fan2021anticipating,gauthier2021next,canaday2021model,flynn2023seeing,kim2023neural,zhai2023emergence, zhai2023model,lin2024learning,zhai2025bridging,SZHL:2025}. 

Transformers have also been applied to dynamical systems, for example in reconstructing unseen dynamics from sparse observations~\cite{zhai2025bridging} and in large-scale zero-shot models for chaotic forecasting~\cite{zhang2024zero,liu2024llms,LBG:2025}. In addition, another related line of work is physics-informed neural networks (PINNs), which incorporate physical constraints into the network~\cite{cuomo2022scientific}, and aligns with the idea of including system parameters as additional input channels to improve model performance. However, the capability of Transformers in optimal output estimation problem in dynamical systems has been questioned~\cite{du2023can}. 

\subsection{Machine learning for critical transitions}

Critical transitions and tipping are abrupt and often irreversible changes in system state that occur once a stability threshold is crossed. Machine learning, in particular reservoir computing, has been introduced to this field, where the model is trained on trajectories from known parameters and then used to infer and predict critical transitions under unseen parameters~\cite{KFGL:2021a}. Shortly thereafter, it was reported~\cite{kim2021teaching} that reservoir computing can interpolate and extrapolate dynamics, effectively learning to infer global temporal structure from local examples. It was demonstrated~\cite{bury2021deep} that deep learning can also serve as an early-warning tool, outperforming traditional early-warning signal methods. Since then, machine learning approaches have been studied systematically as digital twins of dynamical systems for predicting critical transitions: models that evolve in parallel with the physical system, remain synchronized with observational data, and can be probed under different parameter regimes~\cite{KFGL:2021b,kong2023reservoir}. For example, a data-driven framework for predicting tipping in nonautonomous systems was articulated~\cite{PKMZGHL:2024}, with real-world applications. it has been shown that deep learning models can predict rate-induced tipping~\cite{huang2024deep}. In addition, machine learning has been applied to reduce complex models and extract low-dimensional tipping mechanisms~\cite{fabiani2024task}.

\section{Machine-learning methods} \label{appendix:ML_Methods}

\begin{table*} [ht!]
\caption{Reservoir computing hyperparameters}
\label{tab:rc_hyper}
\begin{ruledtabular}
\begin{tabular}{l c c c c}
\textbf{Hyperparameter} & \textbf{Food chain} & \textbf{Power system} & \textbf{Ikeda map} & \textbf{Kuramoto-Sivashinsky} \\ \hline
Reservoir size $n$            & 900  & 800  & 400 & 1000 \\
Link probability $P_{L}$      & 0.004 & 0.313  & 0.708 & 0.450 \\
Spectral radius $\rho$        & 2.3  & 1.6  & 0.17 & 0.89 \\
Input scaling $k_{\text{in}}$ & 3.6  & 2.1  & 2.6 & 0.057\\
Parameter scaling $k_p$       & 0.5  & 1.6  & 0.35 &  -0.052 \\
Parameter bias $p_0$          & $-2.2$ & $-3.1$ & $0.47$ & -185\\
Leakage rate $\alpha$         & 0.30 & 1.0  & 1.0 &  1.0 \\
Regularization $\beta$        & $3\times 10^{-5}$ & $1\times 10^{-4}$ & $1\times 10^{-6}$ & $8\times 10^{-5}$ \\
\end{tabular}
\end{ruledtabular}
\end{table*}

\begin{figure*} [ht!]
\centering
\includegraphics[width=0.8\linewidth]{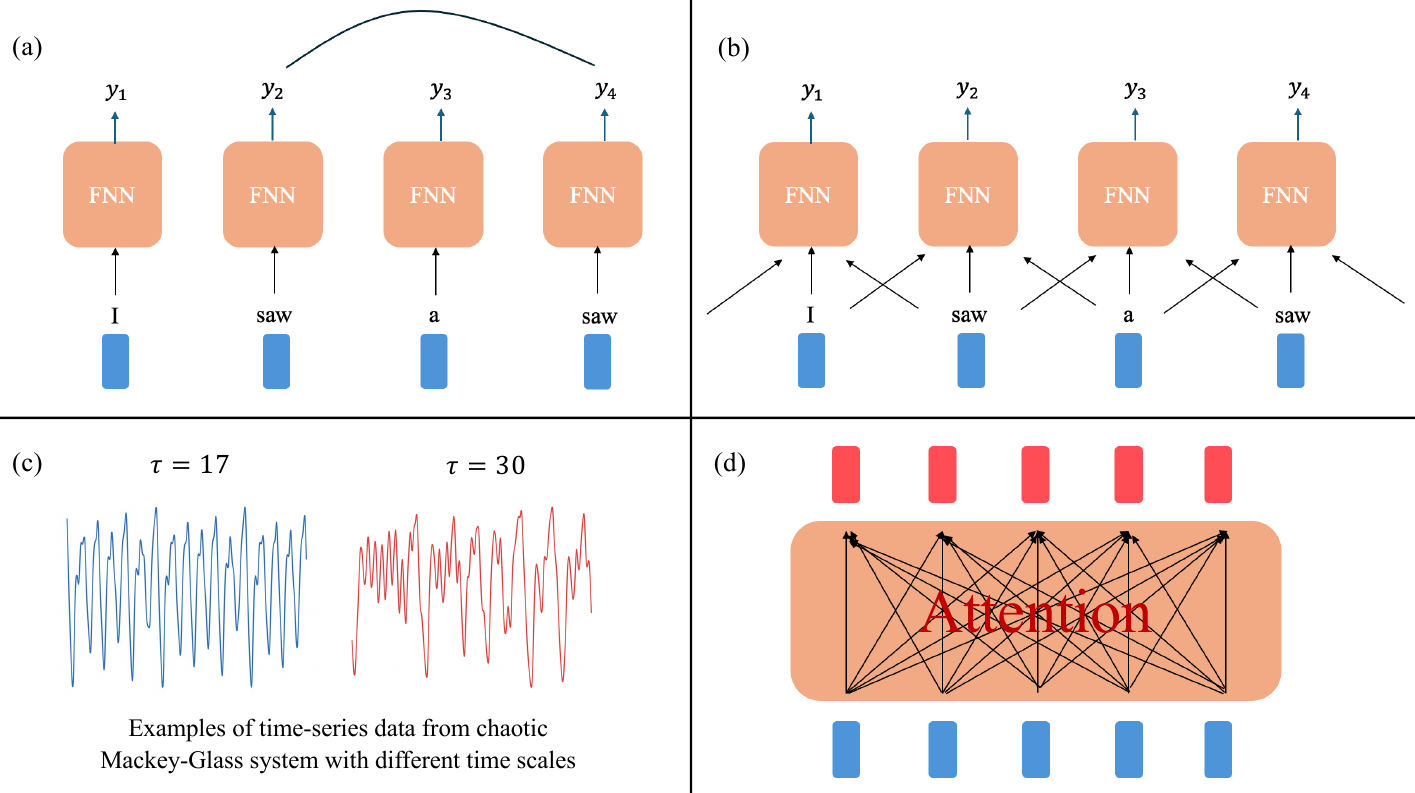}
\caption{\small Conceptual Motivation for the Self-Attention Mechanism. The illustration is based on the limitations of local context windows. (a) The challenge of context-dependent meaning: A model processing words independently cannot distinguish the verb ``saw'' from the noun ``saw'' in the sentence, ``I saw a saw.'' (b) A fixed-size context window provides local information but is an inflexible solution, as the optimal window size is unknown and highly task-dependent. (c) This limitation is analogous to time-series analysis of chaotic systems, where a single, fixed window is inadequate for capturing dynamics across different characteristic timescales (e.g., for different bifurcation parameter values). (d) The Transformer architecture resolves this by employing a global self-attention mechanism, which allows the model to process the entire sequence in parallel and dynamically weigh the relevance of all elements, regardless of their position.}
\label{fig:C20_Beyond_Attention}
\end{figure*}

\subsection{Reservoir computing} \label{appendix:rc}

Reservoir computing is a recurrent neural network framework particularly well suited for modeling dynamical systems. Its main idea is to embed the input time series into a high-dimensional dynamical network with fixed random connections, while only training a linear readout layer. To capture parameter-induced regime changes, we employ parameter-adaptable reservoir computing with an additional input channel for the bifurcation parameter~\cite{KFGL:2021a,KFGL:2021b}. Given an input state $\mathbf{X}(t) \in \mathbb{R}^{d}$ from the target system and a bifurcation parameter $p$, the reservoir state $\mathbf{r}(t) \in \mathbb{R}^{n}$ evolves according to
\begin{eqnarray} \label{eq:rc}
	\mathbf{r}(t+\Delta t) = (1-\alpha) \mathbf{r}(t) + \alpha \tanh ( \mathbb{W} \cdot \mathbf{r}(t) + \nonumber \\ \mathbb{W}_{\text{in}} \cdot \mathbf{X}(t) + \mathbb{W}_p (p+p_0) ),
\end{eqnarray}
where $\mathbb{W} \in \mathcal{R}^{n \times n}$ is the fixed recurrent weight matrix of the reservoir network, the input matrix $\mathbb{W}_{\text{in}} \in \mathcal{R}^{n \times d}$ projects the system input into the reservoir, and the matrix $\mathbb{W}_p \in \mathcal{R}^{n \times 1}$ injects the bifurcation parameter $p$ with parameter bias $p_0$. The leakage rate $\alpha \in (0,1]$ controls the update rate. The nonlinear activation is the hyperbolic tangent.  

The reservoir output is a linear readout of the hidden state:
\begin{align}
	\mathbf{Y}(t) = \mathbb{W}_{\text{out}} \cdot \mathbf{r}(t),
\end{align}
where $\mathbb{W}_{\text{out}} \in \mathcal{R}^{d \times n}$ is trained by ridge regression to minimize
\begin{align}
	\mathcal{L} = \sum_t \| \mathbf{X}(t) - \mathbb{W}_{\text{out}} \cdot \mathbf{r}(t) \|^2 + \beta \|\mathbb{W}_{\text{out}}\|^2,
\end{align}
with $\beta > 0$ a regularization parameter. During training, ground-truth trajectories drive the reservoir; during prediction, the output is fed back, i.e., $\mathbf{X}(t) \leftarrow \mathbf{Y}(t)$. This closed-loop operation turns the trained reservoir computer into a self-evolving dynamical system that, for a given parameter $p$, generates trajectories in the corresponding regime. To maximize the prediction performance, we use Bayesian optimization for determining the values of the hyperparameters during validation, which are listed in Tab.~\ref{tab:rc_hyper}.  

\subsection{Transformer}

\begin{figure*} [ht!]
\centering
\includegraphics[width=0.8\linewidth]{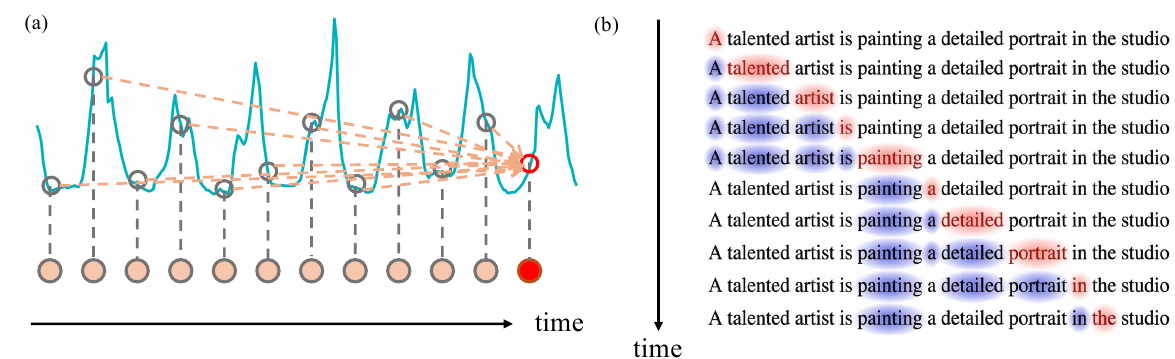}
        \caption{\small Illustration of the ``Full Attention'' Mechanism in Causal or Auto-Regressive Tasks. (a) For time-series forecasting, the model predicts the next value (red point) by attending to all preceding points in the historical data (orange points). This allows it to capture patterns across all available timescales. (b) In language processing, when generating the next word (red), the model calculates attention scores with respect to all previous words in the sequence. The varying intensity of the blue shading schematically represents the attention weights, indicating which prior words the model deems most relevant for predicting the current one.}
\label{fig:C20_Beyond_Attention_2}
\end{figure*}

The Transformer is a deep learning architecture that has become the foundation for most modern natural language processing (NLP) systems, including large language models such as ChatGPT (in fact, the last letter ``T'' in ChatGPT stands for ``Transformer''), as well as tools for machine translation and summarization. The architecture typically consists of two main components: an encoder that processes the input sequence, e.g., a sentence in English, and a decoder that generates the output, e.g., the translated sentence in French. Both components are constructed from layers that rely on attention mechanisms and conventional feed-forward networks. Prior to Transformers, recurrent neural networks such as LSTM were prevalent. These models process text sequentially, which introduces computational bottlenecks and limits their ability to capture long-range dependencies between distant words. The Transformer, introduced in the landmark 2017 paper ``Attention Is All You Need''~\cite{vaswani2017attention}, overcame these limitations. By relying heavily on a mechanism called self-attention, Transformers can process all elements in a sequence in parallel, allowing the model to weigh the importance of all other words when processing a given word. This parallel processing capability has fueled the remarkable success of the Transformer architecture as a foundational model for modern artificial intelligence.

Initially dominant in natural language processing, the Transformer architecture has since been adapted for a wide range of fields, including time-series forecasting~\cite{SZLWZH:2025}, video analysis~\cite{SJENMC:2023}, speech processing~\cite{liu2021tera} and, increasingly, the study of nonlinear dynamics~\cite{zhang2024zero,zhai2025bridging,LBG:2025}. A key advantage of Transformers is their capacity for generalization, particularly in out-of-distribution tasks, a phenomenon exemplified by ``grokking'' in large models~\cite{PBEBM:2022}. In the context of nonlinear dynamics, Transformers have been successfully applied to reconstruct unseen dynamical systems from sparse or incomplete observations~\cite{zhai2025bridging}. Furthermore, large time-series models based on this architecture have been developed for zero-shot forecasting of dynamical systems~\cite{zhang2024zero,LBG:2025}.

\begin{figure*} [ht!]
\centering
\includegraphics[width=0.6\linewidth]{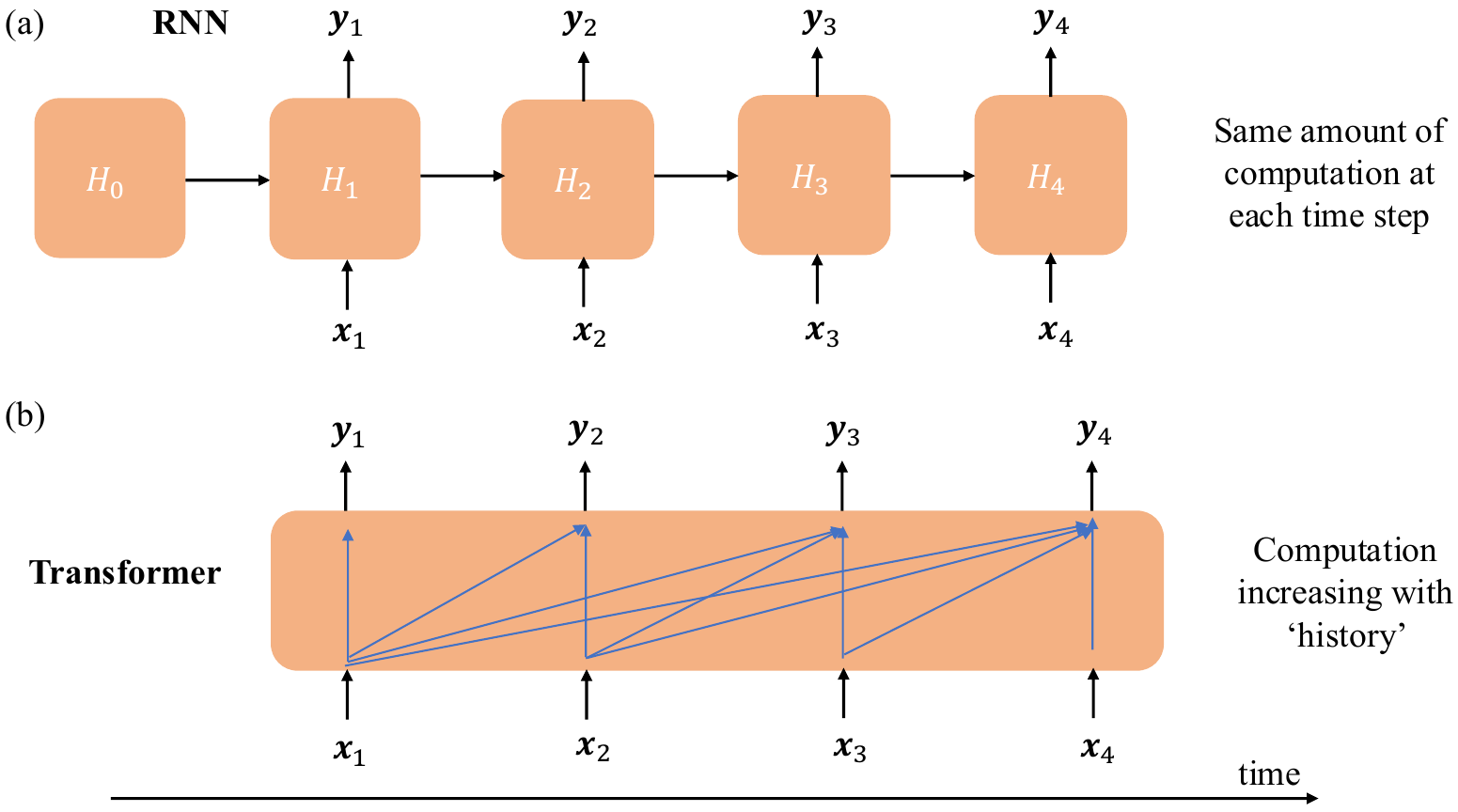}
	\caption{\small Comparison of Sequential Data Processing in Recurrent Neural Networks versus Transformers. (a) A recurrent neural network processes data sequentially. The information from each time step is passed to the next via a hidden state, which acts as the model's memory. This inherently sequential nature creates a computational bottleneck. (b) A Transformer processes the entire sequence in parallel using its self-attention mechanism. This allows it to model dependencies between all elements simultaneously, avoiding the sequential bottleneck and enabling massive parallelization on modern hardware, a key factor in the success of large-scale models.}
\label{fig:C20_Beyond_RNN_Transformer}
\end{figure*}

\subsubsection{Why attention is so important?}

The fundamental importance of the attention mechanism in Transformer can be appreciated by a simple example of language processing. Suppose one wishes to process the sentence ``I saw a saw,'' as illustrated in Fig.~\ref{fig:C20_Beyond_Attention}(a), where an output $y_i$ ($i=1,\ldots,4$) is produced for each word in this sentence and $i$ is the time index. That is, the processing begins with the first word ``I'' at the first time step and proceeds in time to the last word ``saw'' at the fourth time step. The difficulty lies in the second and the fourth words, which are exactly identical but express completely different meanings. Suppose feed-forward networks are used for this task, where one network is trained to produce one word. How can the second and fourth neural networks distinguish the identical word ``saw''? A straightforward approach is to set a ``window'' of inputs for each feed-forward network, allowing it to receive the information not only from the current time step but also from the preceding and following steps, as shown in Fig.~\ref{fig:C20_Beyond_Attention}(b). This leads to the next question: how large should the window size be for different words and for the same word but with different meanings?

The choice of the window, naturally, would depend on the context or the ``surroundings'' of the word. A convenient analogy can be drawn from time-series analysis of nonlinear dynamical systems. Consider, for example, the chaotic Mackey-Glass system described by the time-delayed differential equation~\cite{MG:1977}, where the amount of the time delay $\tau$ is a system parameter. For two different values of $\tau$, the time scales of the oscillations of the dynamical variable can be drastically different, as shown in Fig.~\ref{fig:C20_Beyond_Attention}(c). If a window is used for prediction, its size will be quite different for the two different $\tau$ values. This means that, even for the same dynamical system, the choice of the window for time-series prediction can be different. In language processing, the size of the window for gathering attention for different words will inevitably vary drastically, even for the same word but in the same sentence but with different meanings. The challenge is daunting.

An intuitive solution is to set a large or gigantic window to include all the available information if the goal is, for example, to train a general machine-learning model that can handle time series of the Mackey-Glass system for any value of the time delay. This idea is, in fact, at the heart of the Transformer architecture. In particular, when processing each word (or time point), the Transformer attends to all the other words (or time points) in the sequence, as shown in Fig.~\ref{fig:C20_Beyond_Attention}(d). More concretely, Fig.~\ref{fig:C20_Beyond_Attention_2} presents the ``full-attention'' idea for time series prediction and language processing. For time-series prediction, as shown in Fig.~\ref{fig:C20_Beyond_Attention_2}(a), the model does not need to care about the exact value of $\tau$; it simply takes the entire time series. For language processing, the number of words included in predicting the next word increases with time, as shown in Fig.~\ref{fig:C20_Beyond_Attention_2}(b).

The attention mechanism is what makes Transformers so powerful, for the following reasons. First, it is essential to capturing context. In a sentence, the meaning of a word often depends on other words, sometimes far away. For example, in the sentence ``The book that the professor recommended was fascinating,'' the word ``was'' relates to ``book'', not ``professor''. The attention mechanism helps the model focus on relevant words (like ``book'') when processing another word (``was''). Second, the processing is done in parallel. Unlike recurrent neural networks, Transformers do not read sequences one step at a time. Attention lets them process all words simultaneously, identifying relationships between them in a single step. This makes training much faster and more efficient. Third, it is able to handle long-range dependencies. In long texts, earlier models struggled to remember distant words. The attention mechanism allows every word to directly connect to every other word, regardless of their distance, which helps preserve meaning over long passages. Finally, it offers flexibility and interpretability. Each ``attention head'' can focus on different aspects of meaning: one might track subjects and verbs, another might look for adjectives describing nouns, etc. This makes the model’s internal workings more interpretable than older architectures such as recurrent neural networks.

It is worth noting that the recurrent neural-network architecture, such as reservoir computing, can also handle sequential data, as schematically illustrated in Fig.~\ref{fig:C20_Beyond_RNN_Transformer}(a). In particular, given an initial hidden state $\mathbf{H}_0$ at $t=0$, an input vector $\mathbf{x}_1$ produces an output vector $\mathbf{y}_1$ at $t = 1$ and the hidden state evolves to $\mathbf{H}_1$. For a sequence of input vectors $\{\mathbf{x}_1,\mathbf{x}_2,\ldots\}$, the recurrent neural network produces a sequence of output vectors $\{\mathbf{y}_1,\mathbf{y}_2,\ldots\}$, with the hidden state as memory. This is exactly the training process underlying reservoir computing or LSTM, where the hidden state needs to be computed sequentially. Compared with recurrent neural networks, Transformers require higher computational cost, as they process a larger set of historical data and this set increases with time, as shown in Fig.~\ref{fig:C20_Beyond_RNN_Transformer}(b). However, for large-scale training, Transformers offer a major advantage: the entire sequence can be processed in parallel. This parallelism has enabled the rapid development of today's large language models such as ChatGPT, Claude, and Gemini.

\begin{figure*} [ht!]
\centering
\includegraphics[width=0.7\linewidth]{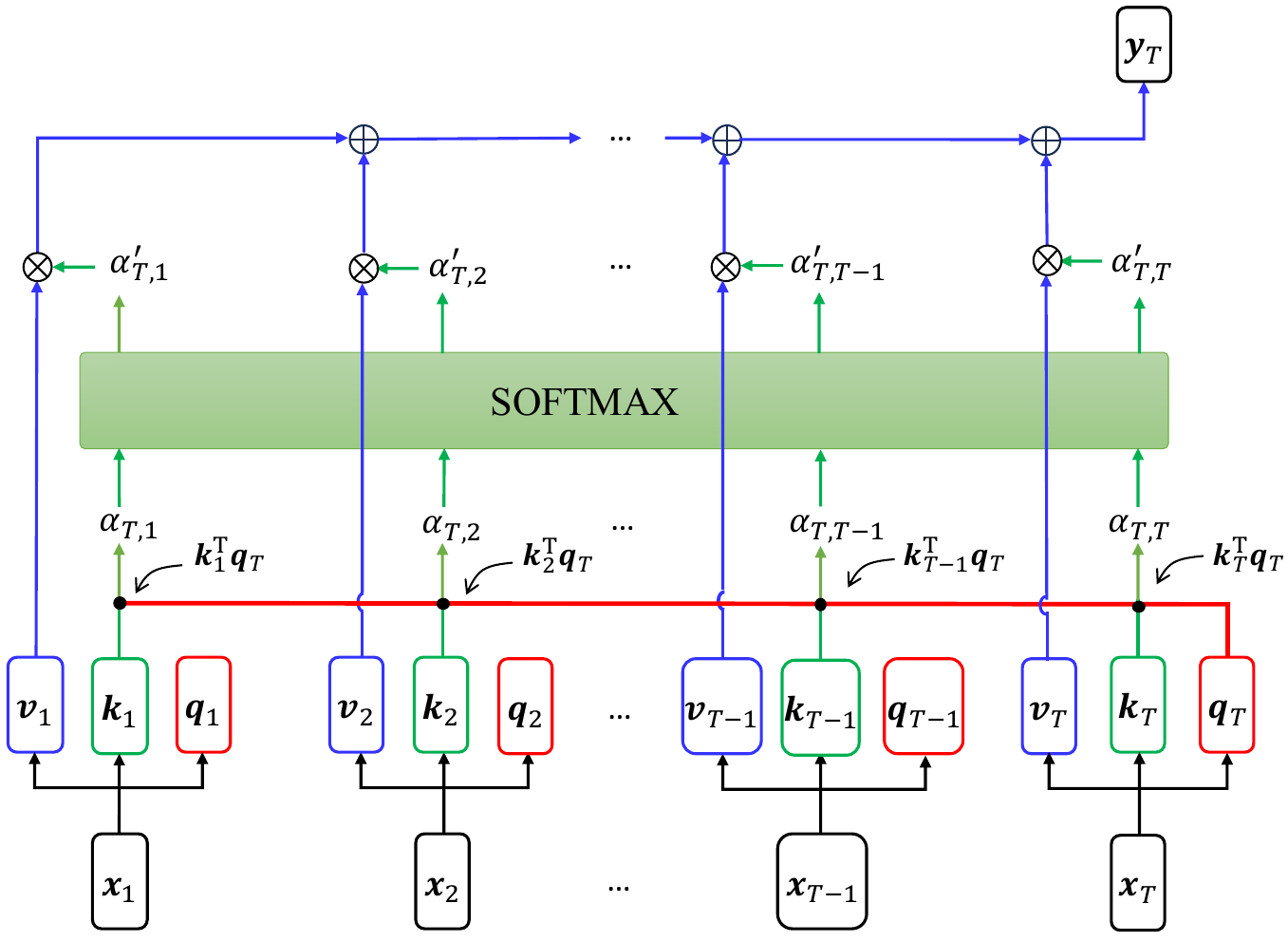}
        \caption{\small Mathematical Architecture of the Scaled Dot-Product Attention Mechanism. The figure illustrates the process for computing a single output vector, $\mathbf{y}_T$, as follows: (i) Score Calculation: The query vector of the current position, $\mathbf{q}_T$, is compared against every key vector ($\mathbf{k}_1, \ldots, \mathbf{k}_T$) in the sequence via a dot product to compute raw attention scores. (ii) Normalization: These scores are normalized using a Softmax function, creating a set of weights that represent the relevance of each input position to the current output. (iii) Weighted Sum: The final output vector, $\mathbf{y}_T$, is computed as a weighted sum of all value vectors ($\mathbf{v}_1, \ldots, \mathbf{v}_T$) using the normalized attention weights. This entire procedure is performed for every position $i$ in the sequence (using query $\mathbf{q}_i$) to generate the full output sequence.}
\label{fig:C20_Beyond_Softmax}
\end{figure*}

\subsubsection{Mathematical Underpinning of Attention Mechanism}

Figure~\ref{fig:C20_Beyond_Softmax} illustrates the basic mathematical steps involved in realizing the attention mechanism. Given an input sequence of $T$ vectors: $\mathbb{X} \equiv \big( \mathbf{x}_1,\mathbf{x}_2,\ldots,\mathbf{x}_T\big)$ up to time $T$, the goal is to generate a sequence of output vectors: $\mathbb{Y} \equiv \big(\mathbf{y}_1,\mathbf{y}_2,\ldots,\mathbf{y}_T\big)$, taking into account the entire dataset $\mathbb{X}$. The dimensions of the input and output vectors can generally be different, denoted as $D_{\rm in}$ and $D_{\rm out}$, respectively. The vector sequences $\mathbb{X}$ and $\mathbb{Y}$ can be interpreted as the input and output data matrices with the respective dimensions $D_{\rm in}\times T$ and $D_{\rm out}\times T$. The central idea underlying Transformer is a trio of three vectors: $\mathbf{q}$, $\mathbf{k}$, and $\mathbf{v}$, named as the query, key, and value vectors, respectively, all with the same dimension $D_{\rm out}$. For an input vector $\mathbf{x}$, the vector triple is generated as
\begin{align} \nonumber
        \mathbf{q} &= \mathbb{W}_q\cdot\mathbf{x}, \\ \label{eq:C20_Beyond_Trio_Vector}
        \mathbf{k} &= \mathbb{W}_k\cdot\mathbf{x}, \\ \nonumber
        \mathbf{v} &= \mathbb{W}_v\cdot\mathbf{x},
\end{align}
where $\mathbb{W}_q$, $\mathbb{W}_k$, and $\mathbb{W}_v$ are $D_{\rm out}\times D_{\rm in}$ matrices, whose elements are determined through training. Performing the operations in Eq.~(\ref{eq:C20_Beyond_Trio_Vector}) at each time step leads to a sequence of $T$ triples: 
\begin{align} \nonumber
	& \{(\mathbf{q}_1,\mathbf{k}_1,\mathbf{v}_1), (\mathbf{q}_2,\mathbf{k}_2,\mathbf{v}_2),\ldots,(\mathbf{q}_{T-1},\mathbf{k}_{T-1},\mathbf{v}_{T-1}), \\ \nonumber
	& \ \ (\mathbf{q}_T,\mathbf{k}_T,\mathbf{v}_T)\}, 
\end{align}
as shown in Fig.~\ref{fig:C20_Beyond_Softmax}.

To obtain the output vector $\mathbf{y}_T$ at time $T$, one uses the value vector $\mathbf{q}_T$ at time $T$ and all the key vectors $\mathbf{k}_i$ ($i=1,\ldots,T$) to generate a set of $T$ scores (scalars):
\begin{align} \label{eq:C20_Beyond_Score_1}
        \alpha_{T,i} = \mathbf{k}_1^{\intercal} \cdot \mathbf{q}_T, \ \ i=1,\ldots,T,
\end{align}
which are kind of initial scores of the input at different times on the output vector at time $T$. These scores can be highly non-uniform, so normalization is needed. This is usually done using the Softmax function, generating
\begin{align} \label{eq:C20_Beyond_Softmax}
        \alpha'_{T,i} = \frac{e^{\alpha_{T,i}}}{\sum_{j=1}^T e^{\alpha_{T,j}}},
\end{align}
where the exponentiation ensures all outputs are positive and all normalized scores sum to one, so the set of $\alpha'_{T,i}$ values can be interpreted as probabilities. In addition, larger input values $\alpha_{T,i}$ get higher scores, but every output depends on all the inputs. The set of normalized scores are then multiplied to their respective value vectors, and the resulting scaled vectors are summed to yield the output vector $\mathbf{y}_T$:
\begin{align} \label{eq:C20_Beyond_Transformer_Output}
        \mathbf{y}_T = \sum_{j=1}^T \alpha'_{T,j} \mathbf{v}_j.
\end{align}
To generate the output at time step $T-1$, one simply replaces the query vector $\mathbf{q}_T$ in Eq.~(\ref{eq:C20_Beyond_Score_1}) by $\mathbf{q}_{T-1}$ and repeats the process in Eqs.~(\ref{eq:C20_Beyond_Softmax}) and (\ref{eq:C20_Beyond_Transformer_Output}). This can be done for all query vectors, leading to the output sequence $\mathbb{Y}$.

The power of Transformer lies in its parallel-computing capability. This can be seen by casting the mathematical procedure in matrix algebra. In particular, from the $T$ query, key, and value vectors, one can define the respective query, key, and value matrices:
\begin{align} \nonumber
\mathbb{Q} &\equiv \big(\mathbf{q}_1,\mathbf{q}_2,\ldots,\mathbf{q}_T\big) = \mathbb{W}_q\cdot \mathbb{X}, \\ \label{eq:C20_Beyond_QKV_Matrices}
\mathbb{K} &\equiv \big(\mathbf{k}_1,\mathbf{k}_2,\ldots,\mathbf{k}_T\big) = \mathbb{W}_k\cdot \mathbb{X}, \\ \nonumber
\mathbb{V} &\equiv \big(\mathbf{v}_1,\mathbf{v}_2,\ldots,\mathbf{v}_T\big) = \mathbb{W}_v\cdot \mathbb{X},
\end{align}
which are all $D_{\rm out} \times T$-dimensional. The set of $T\times T$ scores constitutes a $T\times T$ matrix given by
\begin{align} \label{eq:C20_Beyond_Score_Matrix}
        \mathbb{S} \equiv \mathbb{K}^{\intercal}\cdot \mathbb{Q} = \begin{pmatrix}
                \alpha_{1,1} & \alpha_{1,2} & \ldots & \alpha_{1,T} \\
                \alpha_{2,1} & \alpha_{2,2} & \ldots & \alpha_{2,T} \\
                \vdots       & \vdots       & \vdots & \vdots \\
                \alpha_{T,1} & \alpha_{T,2} & \ldots & \alpha_{T,T}
                           \end{pmatrix}.
\end{align}
The output matrix $\mathbb{Y}$ can then be expressed as
\begin{align} \label{eq:C20_Beyond_Output_Matrix}
\mathbb{Y} = \mathbb{S}' \cdot \mathbb{V}^{\intercal},
%\mathbb{Y} = \mathbb{V} \cdot \mathbb{S}'.
\end{align}
where $\mathbb{S}'$ is the normalized score matrix through a row-based Softmax operation. Computationally, the matrix operations in Eqs.~(\ref{eq:C20_Beyond_QKV_Matrices}), (\ref{eq:C20_Beyond_Score_Matrix}), and (\ref{eq:C20_Beyond_Output_Matrix}) can be carried out in parallel.

\subsubsection{Output of Transformer}

At the output layer, a linear projection maps the final hidden representation back to the original dimension $d$, yielding the next-step prediction $\hat{X}$. Multi-step forecasts are obtained in closed loop by feeding predictions back into the model. To account for parameter-induced behavior shifts, we extend the input with an additional channel containing the bifurcation parameter. The hyperparameters are identical for the three benchmark systems and are listed in Tab.~\ref{tab:Transformer_hyper}.  

\begin{table}[b]
\caption{Transformer hyperparameters}
\label{tab:Transformer_hyper}
\begin{ruledtabular}
\begin{tabular}{l c}
\textbf{Hyperparameter} & \textbf{Value} \\ \hline 
Embedding dimension $N$       & 128 \\
Hidden size (FNN)             & 256 \\
Number of layers $N_b$        & 4 \\
Number of attention heads     & 4 \\
Dropout rate                  & 0.2 \\
Maximum sequence length $L_{\max}$   & 512 \\
Noise level (training)        & 0.05 \\
Total parameters              & $\sim 5\times10^6$ \\
\end{tabular}
\end{ruledtabular}
\end{table}

\subsection{Parameter-adaptable reservoir computing and Transformer} \label{appendix:PARC_Transformer} 

For both Transformer and reservoir computing, we use the same $(d+1)$-dimensional input: the $d$-dimensional state concatenated with a parameter channel. In reservoir computing, the input terms in Eq.~(\ref{eq:rc}) can be written as
\begin{align}
	\mathbb{W}_{\text{in}} \cdot \mathbf{X}(t) + \mathbb{W}_p p = \big[ \mathbb{W}_{\text{in}},\, \mathbb{W}_p \big] \cdot \big[ \mathbf{X}(t), p\big]^{\intercal},
\end{align}
for $p_0 = 0$, demonstrating that the parameter can simply be appended as the last input channel. Here, $\mathbb{W}_{\rm in}$ and $\mathbb{W}_p$ are written separately because they are drawn from different ranges to match the scaling of states and parameter. In the Transformer, the parameter channel is appended in exactly the same way, and its input weights are adjusted during training. In addition, to make the comparison fair, we also include a parameter bias for the Transformer, as $p_0$ in reservoir computing. With this bias, the Transformer model is denoted as Transformer$_b$, with prediction results listed in Table~\ref{tab:performance}.

\section{Additional benchmark systems} \label{appendix:results}

\subsection{Power system} 

\begin{figure*} [t!]
\centering
\includegraphics[width=\linewidth]{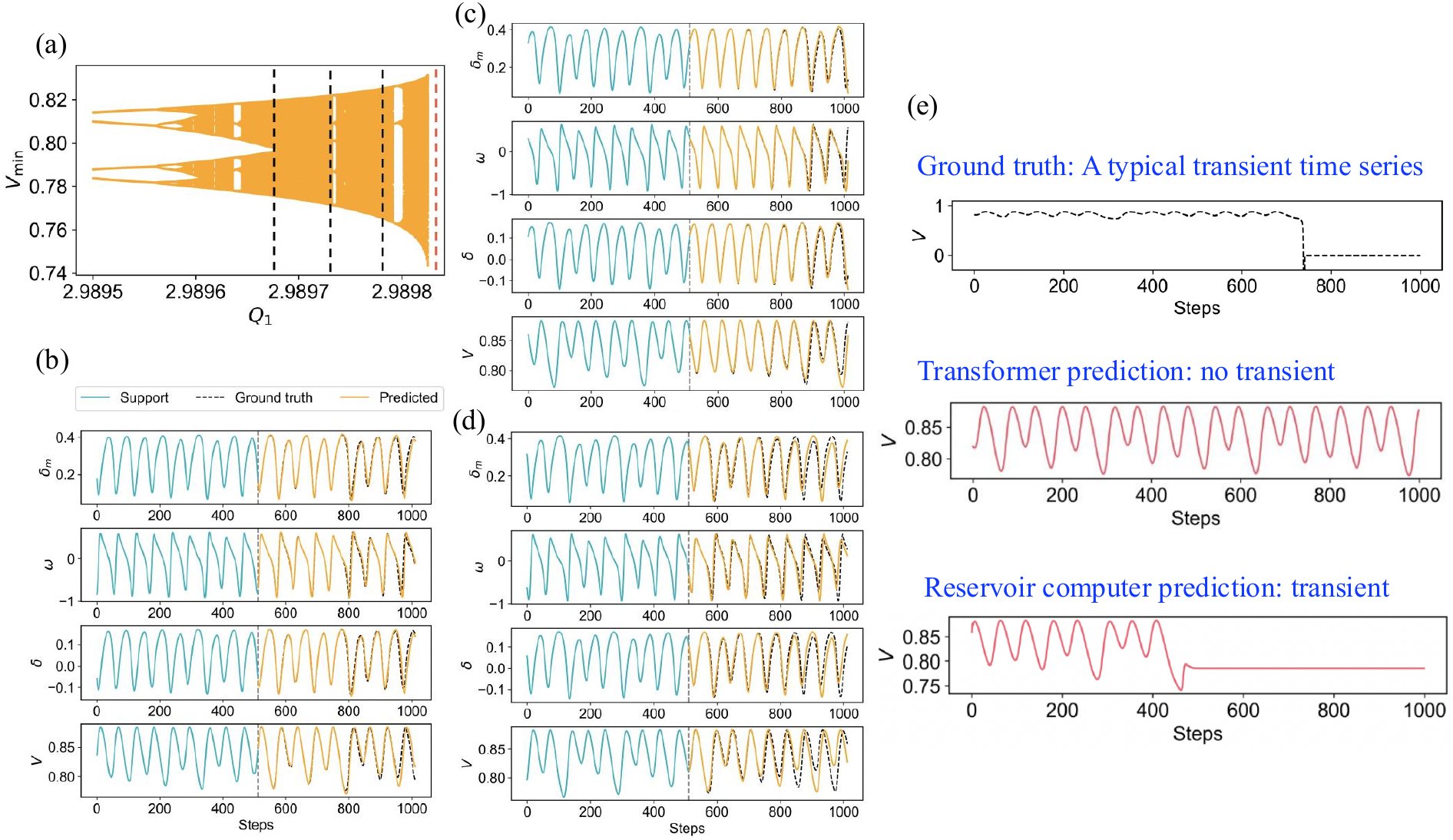}
\caption{Critical transition prediction in the power system. (a) Bifurcation diagram with training parameters (black dashed) and the testing parameter beyond the critical point (red dashed). (b–d) Blue curves indicate warm-up input, and orange curves denote Transformer predictions at the three training parameters $Q_1$. (e) Comparison of Transformer and RC predictions at $Q_1 = 2.989830 > Q_{1c}$. The ground truth trajectory collapses after a transient; reservoir computing reproduces this collapse, whereas the Transformer continues oscillating.}
\label{fig:voltage}
\end{figure*}

\subsubsection{System description}

We consider an electrical power systems including voltage collapse~\cite{dobson1989towards, wang1994bifurcations}. The system exhibits transient chaos prior to collapse, making it an appropriate testbed for critical transition prediction. The model consists of four coupled differential equations, describing the rotor angle $\delta_m$, rotor speed $\omega$, load voltage phase $\delta$, and load voltage magnitude $V$: 
\begin{align} 
\dot{\delta}_m &= \omega, \\ 
M \dot{\omega} &= - d_m \omega + P_m - E_m V Y_m \sin(\delta_m - \delta), \\ \nonumber 
K_{qw} \dot{\delta} &= -K_{qv2} V^2 - K_{qv} V + Q(\delta_m, \delta, V) \\  
	& - Q_0 - Q_1, \\ \nonumber 
T K_{qw} K_{pv} \dot{V} &= K_{pw} K_{qv2} V^2 + (K_{pw} K_{qv} - K_{qw} K_{pv}) V \\ \nonumber
&\quad + K_{qw}[P(\delta_m, \delta, V) - P_0 - P_1] \\ \label{eq:Power_System}
&\quad - K_{pw}[Q(\delta_m, \delta, V) - Q_0 - Q_1], 
\end{align}
where $V \angle \delta$ is the load voltage, $E_m \angle \delta_m$ is the generator terminal voltage, $E_0 \angle 0$ is the infinite bus voltage, and $\omega$ is the rotor angular speed. The load consists of a constant $PQ$ component in parallel with an induction motor. The real and reactive powers supplied to the load are
\begin{align}
P(\delta_m, \delta, V) &= -E_0' V Y_0' \sin \delta + E_m V Y_m \sin(\delta_m - \delta), \\
Q(\delta_m, \delta, V) &= E_0' V Y_0' \cos \delta - (Y_0' + Y_m) V^2 \nonumber \\ &+ E_m V Y_m \cos(\delta_m - \delta),
\end{align}
where $Q_1$ represents the reactive power demand at the load bus and serves as the bifurcation parameter. As $Q_1$ increases, the system undergoes a boundary crisis bifurcation: for $Q_1 < Q_{1c}$, a periodic or chaotic attractor exists; for $Q_1 > Q_{1c}$, trajectories enter transient chaos and eventually collapse, i.e., the load voltage drops precipitously. The critical value is located at $Q_{1c} \approx 2.9898256$.

The parameters are chosen, as follows~\cite{KFGL:2021a}: $M = 0.01464$, $C = 3.5$, $E_m = 1.05$, $Y_0 = 3.33$, $\theta_0 = 0$, $\theta_m = 0$, $K_{pw} = 0.4$, $K_{pv} = 0.3$, $K_{qw} = -0.03$, $K_{qv} = -2.8$, $K_{qv2} = 2.1$, $T = 8.5$, $P_0 = 0.6$, $P_1 = 0.0$, $Q_0 = 1.3$, $E_0 = 1.0$, $Y_m = 5.0$, $P_m = 1.0$, $d_m = 0.05$. The adjusted Th\'evenin equivalents $(E_0', Y_0', \theta_0')$ are defined~\cite{KFGL:2021a} in terms of $E_0$, $Y_0$, and $C$. The bifurcation diagram of the system is illustrated in Fig.~\ref{fig:voltage}(a). The narrow chaotic window makes the system especially sensitive to parameter drift and therefore an ideal benchmark for machine learning based digital twins.

\subsubsection{Predicting a crisis} 

During training, trajectories are generated at $Q_1 = [2.98968, 2.98973, 2.98978]$, all within the safe chaotic regime. Testing is performed at $Q_1 = 2.989830 > Q_{1c}$, where the ground truth shows a short chaotic transient followed by a sudden voltage drop. Figures~\ref{fig:voltage}(b-d) show three typical time series predicted by the Transformer at the three training bifurcation parameter values, respectively, and Fig.~\ref{fig:voltage}(e) compares the Transformer and reservoir-computing predictions. Consistent with the results in the main text, both models perform reliably on the training parameters. Afterward, they are evaluated under a parameter shift beyond the critical point. The Transformer, while accurate on training trajectories, do not exhibit sensitivity to parameter shifts to produce collapse behavior. By contrast, reservoir computing correctly follows the transient oscillations and then reproduces the collapse, in agreement with the ground truth dynamics. 

\subsection{Ikeda map} 

\begin{figure*}[ht!]
\centering
\includegraphics[width=\linewidth]{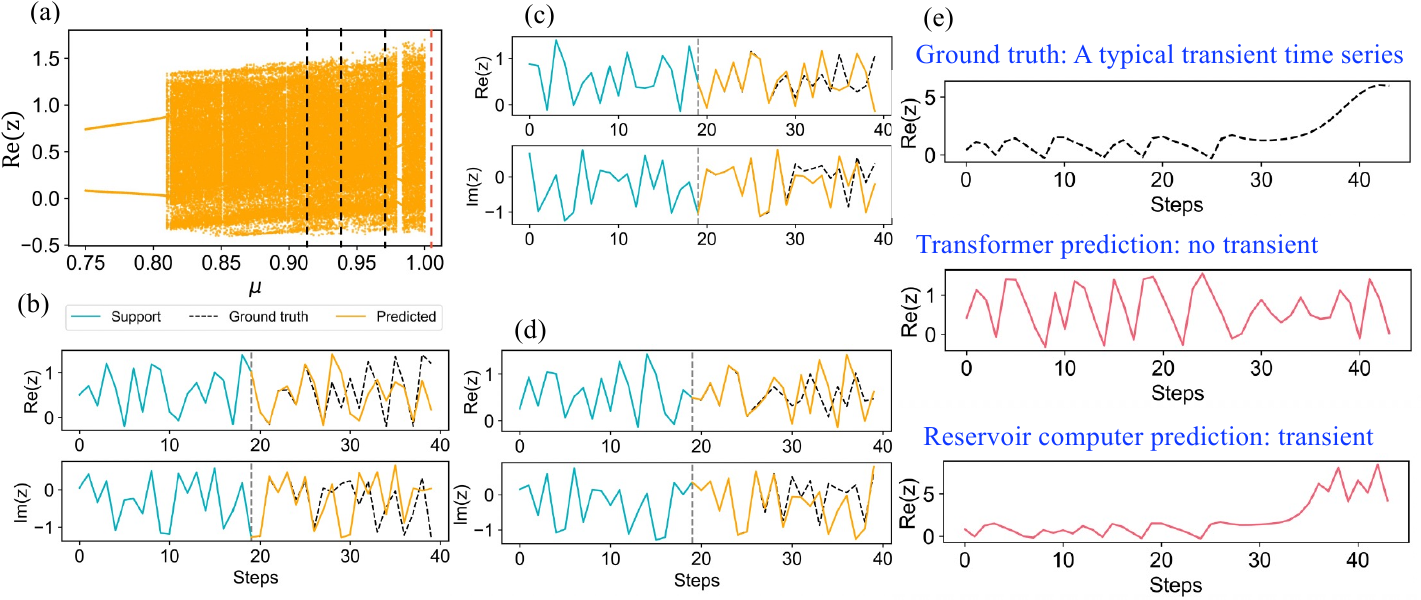}
\caption{Critical transition prediction in the Ikeda map. (a) Bifurcation diagram with training parameters (black dashed) and the testing parameter beyond the critical point (red dashed). (b–d) Blue curves indicate warm-up data, and orange curves denote Transformer predictions at the three training parameters $\mu$. (e) Comparison of Transformer and RC predictions at $\mu = 1.01 > \mu_{c}$. Parameter-adaptable reservoir computing reproduces the collapse after a transient, whereas the Transformer fails and remains oscillatory.}
\label{fig:ikeda}
\end{figure*}

\subsubsection{System description}

The Ikeda map describes the dynamics of a laser pulse propagating through a nonlinear optical cavity~\cite{ikeda1980optical,in1998maintaining}. The map is defined on a complex variable $z \in \mathbb{C}$ and takes the form
\begin{align}
z_{n+1} = \mu + \gamma z_n \exp\!\left( i \kappa - \frac{i \nu}{1 + |z_n|^2} \right), 
\end{align}
where $\mu$ is the dimensionless laser input amplitude, $\gamma$ is the reflectivity coefficient of the cavity mirrors, $\kappa$ is the empty-cavity detuning, and $\nu$ characterizes detuning due to the nonlinear medium. The Ikeda map has long served as a paradigmatic model for studying nonlinear optical dynamics, chaos, and crisis phenomena. A bifurcation diagram is shown in Fig.~\ref{fig:ikeda}(a). Of particular interest is the occurrence of a boundary crisis bifurcation: for $\mu < \mu_c$, the system exhibits periodic or chaotic attractor, while for $\mu > \mu_c$, trajectories display transient chaos before escaping to a constant state, representing collapse. The critical point is located at $\mu_c \approx 1.0027$. In our setting, we follow the parameterization used in~\cite{KFGL:2021b}: $\gamma = 0.9$, $\kappa = 0.4$, and $\nu = 6.0$, while $\mu$ serves as the bifurcation parameter. 

Beyond its relevance to nonlinear optics, the Ikeda map poses a stringent test for equation-discovery methods. Its update law involves a nested non-polynomial nonlinearity, $\exp\!\big(i\kappa - i\nu/(1+|z|^2)\big)$, for which a sparse representation in standard polynomial/Fourier libraries is generally unavailable when only observational time series are given. Consequently, sparse-regression approaches, such as SINDy-type methods, fail to recover a compact governing model, even though the dynamics are low-dimensional. This makes the Ikeda map an important case where black-box digital twins are necessary: reservoir computing succeeds in forecasting the state evolution and predicting the crisis, whereas equation discovery cannot.

% Training is carried out in the chaotic regime at $\mu = 0.91, 0.94, 0.97$, as indicated in Fig.~\ref{fig:ikeda}(d). Testing beyond the crisis, with $\mu > \mu_c$, reveals the transition from sustained chaos to collapse, making the Ikeda map an ideal single-variable benchmark for evaluating machine learning models of critical transitions. 

% The bifurcation diagram is shown in Fig.~\ref{fig:ikeda}(d), where the training parameters are denoted by blue dashed lines and the collapse threshold $\mu_c$ is marked by a black dashed line.

\subsubsection{Predicting a critical transition} 

Training is carried out in the pre-crisis regime at three values of the bifurcation parameter: $\mu = [0.91, 0.94, 0.97]$, while testing is performed for $\mu = 1.01 > \mu_c$, where the ground-truth dynamics shows a collapse characterized by the destruction of the original chaotic attractor with trajectories escaping to a different attractor. Figures~\ref{fig:ikeda}(b-d) present three trajectories generated by the trained Transformer for the three training parameter values, respectively. Figure~\ref{fig:ikeda}(e) shows, for $\mu = 1.01 > \mu_c$, a true trajectory and the trajectories predicted by Transformer and reservoir computing. Again, the Transformer fails to recognize the collapse and instead generates oscillatory trajectories similar to those seen during training. Reservoir computing, in contrast, successfully captures the transient behavior and predicts the eventual collapse. It is particularly noteworthy that the Ikeda map, because its nonlinear update rule cannot be represented in a sparse polynomial or Fourier library, sparse-regression approaches fail. As a result, black-box digital twins are essential for prediction, where reservoir computing proves effective but Transformers again fall short. 

\subsection{Kuramoto-Sivashinsky system} 

\begin{figure*}[ht!]
\centering
\includegraphics[width=\linewidth]{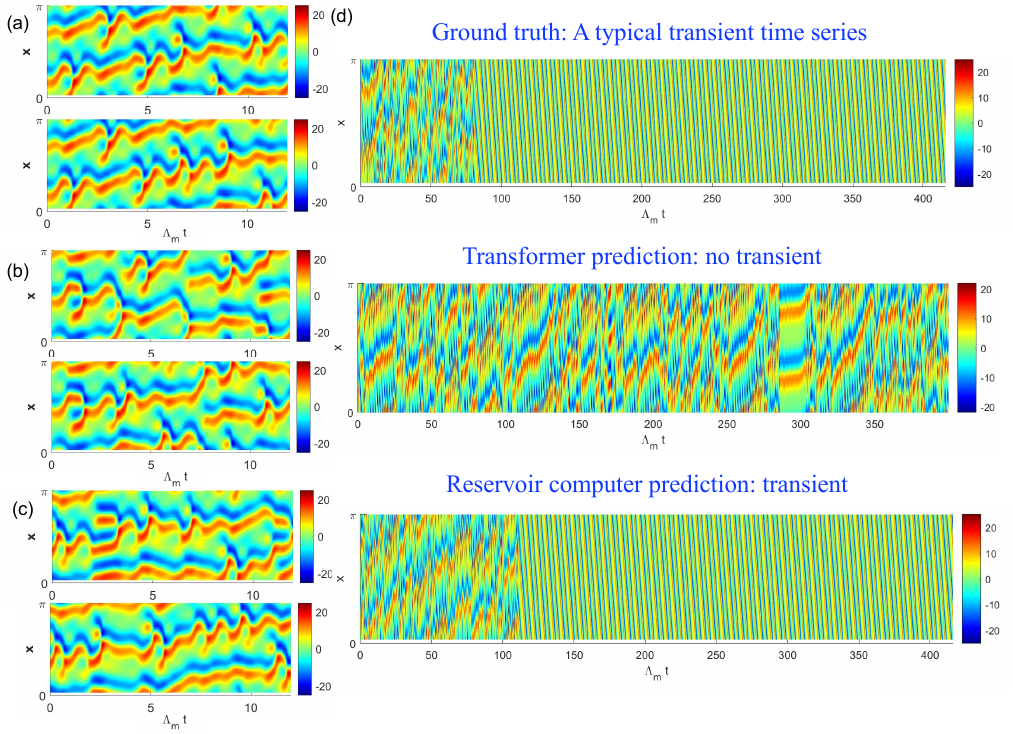}
\caption{Predicting a critical transition in the Kuramoto-Sivashinsky system. (a–c) Ground truth (upper panels) and Transformer predicted spatiotemporal evolution of the KS system for the three training bifurcation-parameter values in the pre-critical regime. (The reservoir-computing predicted results are similar - not shown here.) (d) Comparison of Transformer and reservoir-computing predictions for $\phi = 200.14 > \phi_{c}$, where the former remains in a spatiotemporal chaotic state but the latter correctly generates the critical transition from chaos to a traveling wave.}
\label{fig:ks}
\end{figure*}

\subsubsection{System description}

The Kuramoto-Sivashinsky (KS) system, as described by nonlinear partial differential equation, is a prototypical model for studying nonlinear spatiotemporal dynamics. It was originally derived to model instabilities in laminar flame fronts, and later applied to other physical systems, such as trapped-ion instability in plasmas~\cite{laquey1975nonlinear}. Depending on the system parameters, its solutions range from regular traveling waves to sustained chaos and transient chaotic behavior, which makes the KS equation an informative benchmark for evaluating whether a machine learning model can capture qualitative changes in dynamics when a bifurcation parameter is varied. In this work, we consider the one-dimensional KS equation on a periodic domain
$x \in [0,\pi]$:
\begin{equation}
   \frac{\partial u}{\partial t} + \nu \frac{\partial^4 u}{\partial x^4} + \phi \left(\frac{\partial^2 u}{\partial x^2} + u \frac{\partial u}{\partial x} \right) = 0,
\label{eq:ks}
\end{equation}
where $u(x,t)$ is a scalar field and $\nu$, $\phi$ are system parameters. We fix $\nu = 4$ following standard practice for generating transient chaos and treat $\phi$ as the bifurcation parameter~\cite{KFGL:2021a}.

For $\phi$ below a critical value $\phi_c \approx 200.04$, the system exhibits sustained spatiotemporal chaos characterized by irregular fluctuations in both space and time. When $\phi$ is increased past $\phi_c$, the chaotic attractor is destroyed through a crisis-like transition: trajectories enter a regime of transient chaos and  eventually collapse onto a stable traveling-wave solution. This collapse state is never seen during training, since all training trajectories are generated at parameter values strictly below $\phi_c$.

\subsubsection{Predicting a critical transition} 

A boundary crisis occurs at $\phi_c \approx 200.04$. Training is performed in the safe regime at three parameter values $\phi \in \{196, 197, 198\}$, where all trajectories remain chaotic. Testing is conducted at a supercritical parameter $\phi = 200.14 > \phi_c$, at which the ground truth collapses after a transient. For each of the three training bifurcation-parameter values in the pre-critical regime, both Transformer and reservoir computing can accurately reproduce the chaotic dynamics, as exemplified in Figs.~\ref{fig:ks}(a-c). For $\phi = 200.14$ in the post-critical regime, however, the two models behave quite differently. As shown in Fig.~\ref{fig:ks}(d), reservoir computing successfully captures the transient chaotic phase and the subsequent collapse into a traveling-wave state, whereas the Transformer fails and continues to generate chaotic predictions indefinitely.

\begin{table*}[ht!]
\caption{Performance on critical transition prediction on Kuramoto-Sivashinsky system}
\label{tab:performance_ks}
\begin{ruledtabular}
\begin{tabular}{l l c c c c c}
% \toprule
\textbf{system} & \textbf{model} & 
$P_c(\text{collapse})\uparrow$ & \textbf{pred.\ crit.} & \textbf{gt crit.} & \textbf{$T_{\text train}$[s]}$\downarrow$ \\
KS & Transformer  & 0.30  & 200.332  & 200.04  & 8500 \\
    & Reservoir computing  & 0.94  & 200.163  & 200.04   & 2.29 \\
% \bottomrule
\end{tabular}
\end{ruledtabular}
\end{table*}

It is important to note that the collapse behavior in the KS system differs from the other low-dimensional benchmark systems where collapse corresponds to a state variable abruptly falling to a fixed point far away from the original chaotic attractor. In contrast, the KS system transitions from broadband spatiotemporal chaos to a nearly monochromatic traveling-wave solution. To quantify this transition, we analyze the temporal signal extracted at a representative spatial location (e.g., the midpoint of the domain). For each sliding time window of length $L_w$, we denote the windowed segment by $u(t)$ and compute its discrete Fourier transform, $U_k = \mathrm{FFT}(u(t))$. The corresponding power spectrum is $P_k = |U_k|^2$. We then define the spectral concentration ratio
\begin{equation} \label{eq:ks_ratio}
    R = \frac{\max_{k>0} P_k}{\sum_{k>0} P_k},
\end{equation}
where the sum and maximum are taken over all nonzero positive frequencies. In the chaotic regime, the spectrum is broadband and $R$ is relatively small. After collapse, the traveling-wave solution exhibits a more concentrated spectrum, and $R$ becomes significantly larger.

Let $R_i$ denote the ratio computed from the $i$-th sliding window, and let $\widetilde{R}_i$ be a short moving-average smoothing of $\{ R_i \}$. A collapse is declared if
\begin{equation}
    \widetilde{R}_i > \theta_R \quad \text{for at least $m$ consecutive windows},
\end{equation}
where we set the threshold to $\theta_R = 0.5$. To avoid false positives, we additionally require that the majority of windows prior to the transition satisfy $\widetilde{R}_j < \theta_R$, ensuring that the trajectory was chaotic before becoming spectrally concentrated. The collapse time is then defined as the center time of the first window fulfilling these conditions.

Table~\ref{tab:performance_ks} summarizes the statistical results. Both reservoir computing and Transformer require longer training time on this high-dimensional spatiotemporal system. The Transformer performs poorly even after extending the sequence length to $L_{\max}=1024$ and downsampling the data (every two points) to fit memory constraints. Reservoir computing achieves a collapse prediction rate of $94\%$, which is lower than that for the low-dimensional benchmarks but remains substantially higher than that of the Transformer.

%\subsection{Transformer} \label{appendix:Transformer} 

%We adopt a decoder-only Transformer architecture with a causal attention mask, so that each state depends only on its past. The input multivariate time series $X \in \mathbb{R}^{L \times d}$, where $L$ is the sequence length and $d$ the variable dimension, is first projected through a linear layer to an embedding dimension $N$. To preserve temporal ordering, sinusoidal positional encodings are added to the embeddings before entering the Transformer blocks. Each Transformer block processes a hidden representation $H^{(\ell)} \in \mathbb{R}^{L \times N}$, where $\ell$ denotes the layer index. Within each block, queries, keys, and values are computed as
%\begin{align}
%Q^{(\ell)} = H^{(\ell)} W_Q^{(\ell)}, \quad 
%K^{(\ell)} = H^{(\ell)} W_K^{(\ell)}, \quad 
%V^{(\ell)} = H^{(\ell)} W_V^{(\ell)},
%\end{align}
%and the scaled dot-product attention is
%\begin{align}
%\text{Attention}(Q^{(\ell)},K^{(\ell)},V^{(\ell)}) = \text{softmax}\!\left(\frac{Q^{(\ell)}K^{(\ell)\top}}{\sqrt{d_k}}\right)V^{(\ell)}.
%\end{align}
%Multiple heads are computed in parallel, concatenated, and linearly projected, after which residual connections, layer normalization, and a position-wise feed-forward network with ReLU activation are applied. The resulting $H^{(\ell+1)}$ serves as the input to the next block. Stacking $N_b$ such blocks enables the model to capture long-range temporal dependencies.  

\section{Further experiments on Transformers} \label{appendix:further_transformer}

\begin{figure*}[ht!]
\centering
\includegraphics[width=\linewidth]{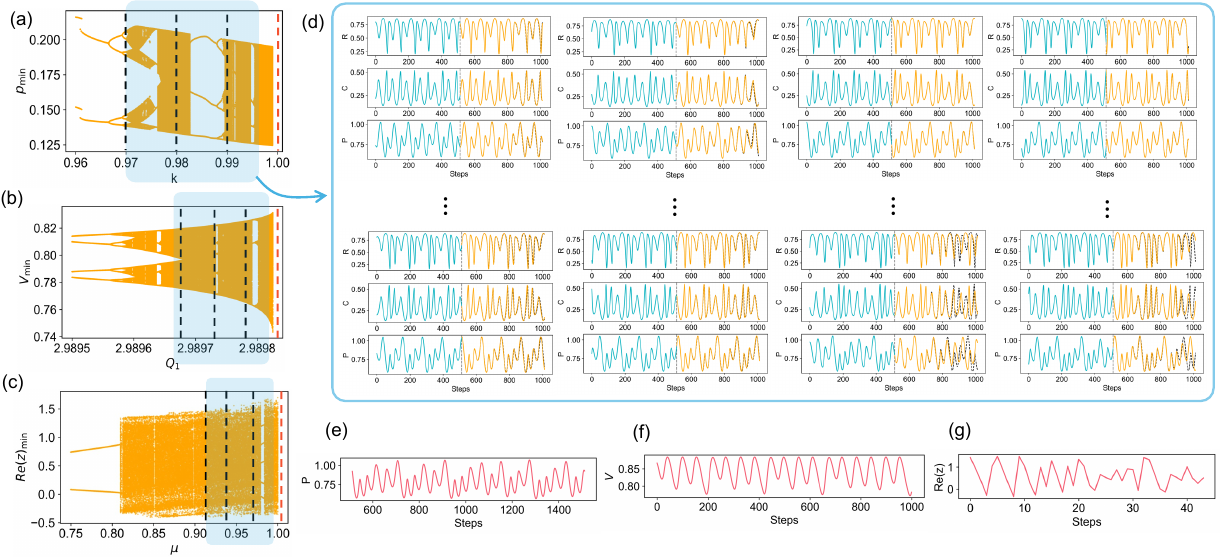}
\caption{Effect of increased training data on Transformer. (a-c) Expanded training parameter ranges (blue shaded), with parameters used in the main text (black dashed) and the supercritical test parameter (red dashed). (d) Example Transformer predictions for the food-chain system, trained with 27 parameter values. (e-g) Transformer critical transition predictions for the food-chain system, power system, and Ikeda map, respectively. Although the model performs well across many safe parameters, it continues to generate oscillations rather than collapse beyond the critical point.}
\label{fig:more_bifur}
\end{figure*}

\begin{figure*}[ht!]
\centering
\includegraphics[width=0.8\linewidth]{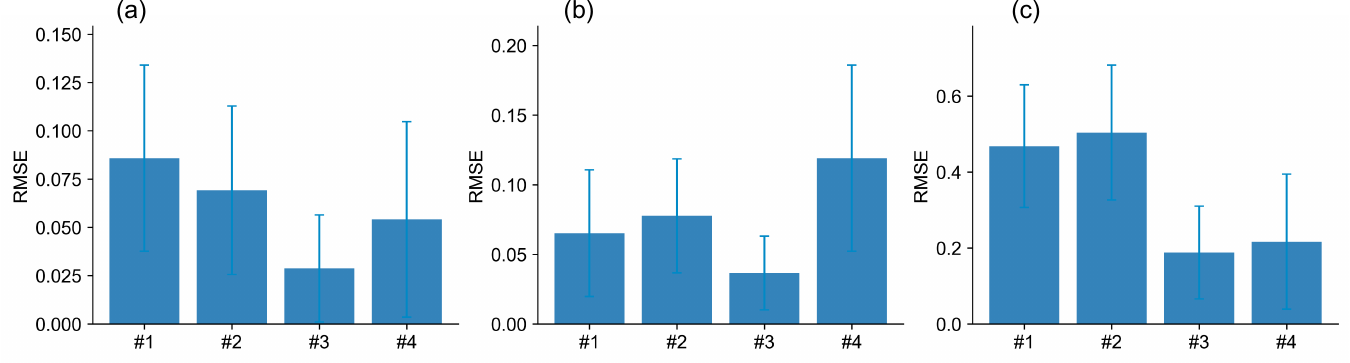}
\caption{Multi-step forecasting accuracy of different Transformer configurations. Error bars denote standard deviations across 10 independent realizations. Model \#3 corresponds to the baseline used in the main text.}
\label{fig:rmse}
\end{figure*}

Transformers are well known to be data-hungry and prone to overfitting. In our main experiments, we demonstrated that for reservoir computing, training on trajectories from only three bifurcation parameters was already sufficient to infer unseen regimes with collapse. For fairness, the same training setting was adopted for Transformers. However, with such limited data, Transformers, owing to their quite large number of trainable parameters, may simply memorize the trajectories associated with each training parameter, rather than learning the underlying dynamical rules. This raises the concern that their apparent accuracy in the training regime may reflect overfitting rather than genuine generalization.

To address this concern, we perform additional experiments with denser bifurcation parameter sampling. For each benchmark system, instead of using only three bifurcation parameters, we generated much larger training sets: 27, 21, and 31 parameter values for the food-chain system, power system, and Ikeda map, respectively. These cover the similar parameter ranges as in the main text. The shaded blue intervals in Figs.~\ref{fig:more_bifur}(a–c) denote the expanded training ranges, with black dashed lines indicating the parameters used previously and the red dashed line marking the test parameter beyond the critical point.

Figure~\ref{fig:more_bifur}(d) shows representative Transformer results for the food-chain system. Across a wide variety of training parameters, the model achieves accurate multi-step forecasts, confirming that the network is well trained. Panels (e–g) further illustrate Transformer predictions at supercritical parameters for the food-chain system, power system, and Ikeda map, respectively. In all cases, although the models are well trained in the safe regimes, they continue to produce oscillatory trajectories instead of reproducing collapse. In other words, even with substantially more training data, Transformer performance on critical transition prediction does not improve in any significant way. By contrast, reservoir computing achieves high success rates in collapse prediction using only three training parameters, without requiring such extensive data.

In addition, to further study Transformer performance, we conduct experiments under multiple architectural configurations. Specifically, we vary the embedding dimension, hidden size, number of layers, and number of attention heads. Three representative configurations were selected as [128, 64, 2, 2], [256, 256, 4, 4] and [512, 256, 8, 8], corresponding to models with approximately $6.7\times 10^5$, $2.6\times 10^6$, and $1.5\times 10^7$ trainable parameters, respectively. For simplicity, we denote these as Models \#1, \#2, and \#4. Model \#3 refers to the architecture used in the main text (see Tab.~\ref{tab:Transformer_hyper}). Figure~\ref{fig:rmse} demonstrates multi-step forecasting results averaged over 10 independent realizations. Standard deviations are relatively large, reflecting both the variability across independently trained models and differences across training bifurcation parameters within the same model. Among the four configurations, Model \#3 yields the best multi-step forecasting performance and is therefore adopted as our main Transformer baseline. Moreover, additional tests are conducted, which confirm that none of the alternative configurations improved performance on critical transition prediction.

\section{Additional statistics of critical transition prediction} \label{appendix:add_statistics}

\begin{figure*} [ht!]
\centering
\includegraphics[width=0.8\linewidth]{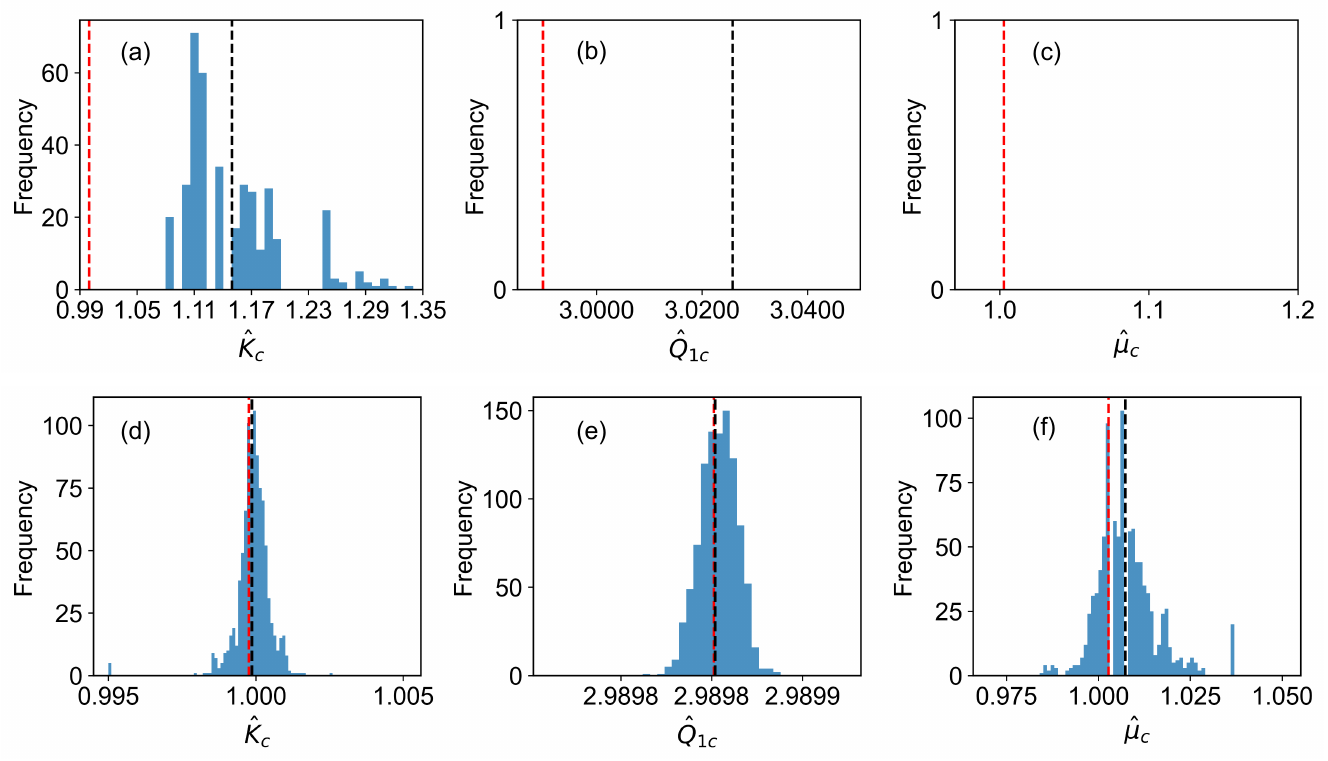}
\caption{Histograms of predicted critical points $\hat{p}_c$. Rows correspond to models (top: Transformer, bottom: reservoir computing). Columns correspond to systems: (a,d) Food-chain system. (b,e) Power system, and (c,f) Ikeda map. Red dashed line denote the ground-truth critical points, where $K_c= 0.99976, Q_{1c}=2.9898256$, and $\mu_c= 1.0027$. Black dashed line indicate the mean predicted critical points: $\langle \hat{K}_c \rangle=1.14958$ and $\langle \hat{Q}_{1c}\rangle=3.02578$ for the Transformer, and $\langle \hat{K}_c \rangle=0.99986$, $\langle \hat{Q}_{1c}\rangle=2.98983$ and $\langle \hat{\mu}_c\rangle=1.0073$ for RC.}
\label{fig:hist}
\end{figure*}

In the main text, we summarized the statistical performance of Transformer and reservoir computing in critical transition prediction in Tab.~\ref{tab:performance}. To provide a more detailed view, we depict histograms of the predicted critical points across all trials, as shown in Fig.~\ref{fig:hist}. For each system and each machine-learning model, we conducted $50$ independent trainings, with $20$ predictions per training (different warm-up segments), yielding $1000$ realizations in total. Specifically, the histograms in Fig.~\ref{fig:hist} show the distribution of predicted critical values $\hat{p}_c$ compared with the ground-truth $p_c$. The first row corresponds to Transformers and the second to reservoir computing. The three columns represent the food-chain system, power system, and Ikeda map, respectively. In each panel, the red dashed line indicates the ground-truth $p_c$, while the black dashed line marks the mean of the predicted values across all realizations. For the Transformer, in the power system only a single collapse prediction was observed among 1000 runs, we take that as the mean and depict as dashed black line in panel (b). For the Ikeda map, no collapse was predicted in any realization, and therefore only the ground-truth line is shown in panel (c). The horizontal axis in each histogram matches the parameter search range used in the experiments; values outside this range are omitted as they lie far beyond the true critical point.

These results further highlight the contrast between the two machine-learning models. For Transformers, collapse predictions are nearly absent in the power system and Ikeda map. Even for the food-chain system, where collapse was predicted in $38\%$ of realizations, the averaged $\hat{p}_c$ values were significantly biased upward relative to $p_c$. In contrast, reservoir computing achieves robust and reliable results: not only did it consistently predict collapse in all realizations, but the predicted critical points were tightly concentrated around the ground truth. 

\begin{figure}[ht!]
\centering
\includegraphics[width=\linewidth]{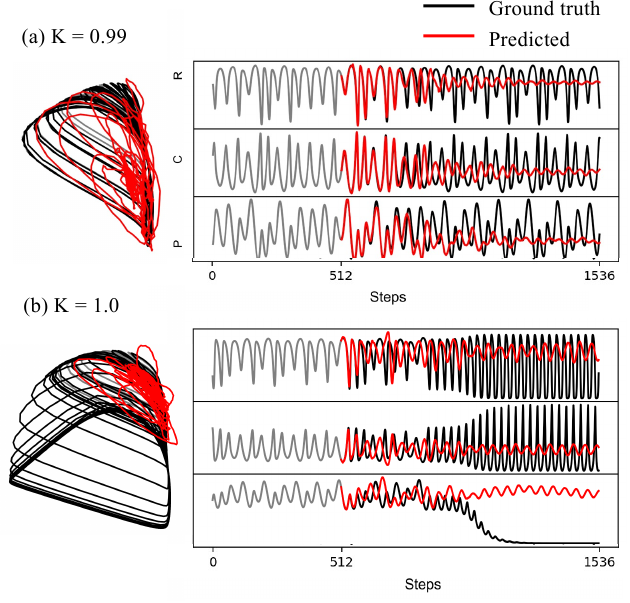}
\caption{Food-chain system predictions by PANDA. (a) Long- (left) and short-term (right) predictions under safe regime. (b) Long- (left) and short-term (right) predictions under collapse. PANDA provides accurate zero-shot predictions for several cycles but fails to maintain long-term accuracy.}
\label{fig:panda}
\end{figure}

\begin{figure}[ht!]
\centering
\includegraphics[width=\linewidth]{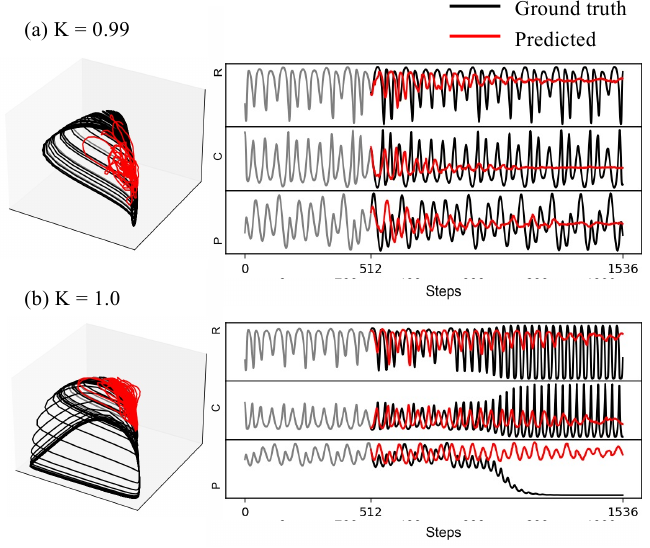}
\caption{Food-chain system predictions by Chronos-2. (a) Long- (left) and short-term (right) predictions under safe regime. (b) Long- (left) and short-term (right) predictions under collapse. Chronos-2 shows limited short-term accuracy and fails in long-term forecasting.}
\label{fig:chronos}
\end{figure}

\section{Failure of two alternative foundations models on system collapse prediction}\label{appendix:foundation}

Recent advances in time series modeling have produced pretrained foundation models that demonstrate strong performance across forecasting, anomaly detection, and imputation tasks~\cite{LBG:2025,ansari2024chronos,das2024decoder,liang2024foundation}. Of particular relevance to nonlinear dynamics are models trained directly on large collections of chaotic systems, which aim to generalize across dynamical regimes in a zero-shot manner~\cite{LBG:2025}. Related efforts have also explored zero-shot learning for dynamical systems reconstruction from sparse observations~\cite{zhai2025bridging}, although the focus there differs from the prediction tasks considered in this study.
In this section, we evaluate two representative models on the problem of system collapse prediction: PANDA, a pretrained model for chaotic dynamics, and Chronos-2, a multivariate foundation model for time-series forecasting.

\subsection{PANDA}

PANDA is a pretrained forecast model specifically designed for nonlinear and chaotic dynamics~\cite{LBG:2025}. It is trained on an extensive synthetic dataset of more than $2\times10^4$ chaotic systems described by ordinary differential equations, generated through evolutionary recombination of known dynamical systems. PANDA utilizes a patched attention architecture, enriched with dynamics-motivated embeddings, such as polynomial and Fourier features. PANDA demonstrates impressive zero-shot forecasting capabilities on unseen chaotic systems, even those described by partial differential equations.

However, its predictions can deteriorate in long-term chaotic attractor reconstructions, where it could drift toward fixed points. Because system collapse prediction requires accurate long-horizon forecasting within the safe regime as a prerequisite, such drift limits its suitability for our task. Figure~\ref{fig:panda} illustrates representative examples of PANDA zero-shot predictions for the food-chain system. PANDA predicts several cycles with high accuracy in both safe and collapse regimes, but cannot reconstruct attractors or detect collapse reliably in the long term.

\subsection{Chronos-2}

Chronos-2~\cite{fatir2025chronos} is the recently released multivariate extension of the Chronos family, designed as a general foundation model for time series forecasting. It incorporates cross-channel attention and can process multivariate trajectories, making it applicable to our tasks. The model is trained using large-scale probabilistic objectives on diverse real-world datasets, and it achieves strong short-term predictive performance in many conventional forecasting tasks.

For chaotic dynamics, Chronos-2 shows reasonable performance, but worse than PANDA in short-term prediction, which is expected given that PANDA is specifically trained on chaotic systems. Furthermore, like other time series foundation models, it fails to preserve attractor geometry during long predictions. As shown in Fig.~\ref{fig:chronos}, Chronos-2 cannot reconstruct chaotic dynamics in the safe regime and does not anticipate parameter-induced collapse. 

\bibliography{Transformer}

\end{document}